\newcommand{\PaperTitle}{The Age of DDoScovery: An Empirical Comparison of \\ Industry and Academic DDoS Assessments}
\newcommand{\ShortPaperTitle}{An Empirical Comparison of Industry and Academic DDoS Assessments}

\documentclass[sigconf]{acmart}

\usepackage[utf8]{inputenc}

\usepackage{hyperref}

\usepackage{listings}
\usepackage{balance}

\usepackage{tikz}
\usepackage{xcolor}
\usepackage{color}
\usepackage{pgfplots}

\usepackage{sistyle}
\SIthousandsep{,}

\usepackage{makecell}
\usepackage{multirow}
\usepackage{booktabs}

\usepackage{amsmath}

\usepackage{comment}
\usepackage{caption}
\usepackage[labelformat=simple]{subcaption}

\usepackage{dblfloatfix} 
\graphicspath{{./figs/}}

\usetikzlibrary{
	arrows.meta,
	calc,
	colorbrewer,
	external,  
	fit,
	matrix,
	positioning,
	shapes,
	shapes.misc,
	backgrounds,
	mindmap
}

\usepackage{url}

\usepackage{hyperref}
\makeatletter
\def\sectionautorefname{\S\@gobble}
\def\subsectionautorefname{\S\@gobble}
\def\subsubsectionautorefname{\S\@gobble}
\makeatother

\usepackage{graphicx,calc}
\newlength\myheight
\newlength\mydepth
\settototalheight\myheight{Xygp}
\newcommand*\inlinegraphics[1]{%
  \settototalheight\myheight{Xygp}%
  \settodepth\mydepth{Xygp}%
  \raisebox{-\mydepth}{\includegraphics[height=\myheight]{#1}}%
}

\usepackage[absolute,showboxes]{textpos}

\copyrightyear{2024}
\acmYear{2024}
\setcopyright{rightsretained}
\acmConference[IMC '24]{Proceedings of the 2024 ACM Internet Measurement Conference}{November 4--6, 2024}{Madrid, Spain}
\acmBooktitle{Proceedings of the 2024 ACM Internet Measurement Conference (IMC '24), November 4--6, 2024, Madrid, Spain}
\acmDOI{10.1145/3646547.3688451}
\acmISBN{979-8-4007-0592-2/24/11}



\newcommand{\result}[1]{}

\definecolor{myred}{cmyk}{0, 0.7808, 0.4429, 0.1412}

\newcommand{\done}[1]{}

\definecolor{positive}{HTML}{3C8031}
\definecolor{negative}{HTML}{AF3235} 
\definecolor{neutral}{HTML}{E1A205} 

\usepackage{pifont}
\newcommand{\upwardTrend}{\textcolor{positive}{\ding{115}}}
\newcommand{\downwardTrend}{\textcolor{negative}{\ding{116}}}
\newcommand{\neutralTrend}{\textcolor{neutral}{\ding{117}}}

\usepackage{xspace}

\newcommand{\eg}{\textit{e.g.,}~}
\newcommand{\ie}{\textit{i.e.,}~}

\newcommand{\one}{({\em i})\xspace}
\newcommand{\two}{({\em ii})\xspace}
\newcommand{\three}{({\em iii})\xspace}
\newcommand{\four}{({\em iv})\xspace}
\newcommand{\five}{({\em v})\xspace}

\newcommand*\circled[1]{\tikz[baseline=(char.base)]{
            \node[shape=circle,draw,inner sep=0.9pt] (char) {\sf \small #1};}}

\makeatletter
\renewcommand{\paragraph}[1]{\vspace*{0.03in}\noindent{\bf #1.}\hspace{0.25ex \@plus1ex \@minus.2ex}}
\newcommand{\paragraphNoDot}[1]{\vspace*{0.03in}\noindent{\bf #1}\hspace{0.25ex \@plus1ex \@minus.2ex}}
\makeatother

\newcommand{\telescope}{\inlinegraphics{emojis/telescope.png}}
\newcommand{\dpi}{\inlinegraphics{emojis/dpi.png}}
\newcommand{\honeypot}{\inlinegraphics{emojis/honeypot.png}}
\newcommand{\scrubbing}{\inlinegraphics{emojis/scrubbing.png}}
\newcommand{\blackhole}{\inlinegraphics{emojis/blackhole.png}}

\makeatletter
\gdef\@copyrightpermission{
  \begin{minipage}{0.3\columnwidth}
   \href{https://creativecommons.org/licenses/by/4.0/}{\includegraphics[width
=0.90\textwidth]{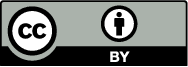}}
  \end{minipage}\hfill
  \begin{minipage}{0.7\columnwidth}
   \href{https://creativecommons.org/licenses/by/4.0/}{This work is licensed
under a Creative Commons Attribution International 4.0 License.}
  \end{minipage}
  \vspace{5pt}
}
\makeatother

\begin{document}

\title[\ShortPaperTitle]{\PaperTitle}

\author{Raphael Hiesgen} 
\affiliation{\institution{HAW Hamburg}
  \city{Hamburg}
  \country{Germany}
}

\author{Marcin Nawrocki} 
\affiliation{\institution{NETSCOUT}
  \city{Westford}
  \state{MA}
  \country{USA}
}

\author{Marinho Barcellos} 
\affiliation{\institution{U of Waikato}
  \city{Hamilton}
  \country{New Zealand}
}

\author{Daniel Kopp} 
\affiliation{\institution{DE-CIX}
  \city{Frankfurt am Main}
  \country{Germany}
}

\author{Oliver Hohlfeld} 
\affiliation{\institution{University of Kassel}
  \city{Kassel}
  \country{Germany}
}

\author{Echo Chan} 
\affiliation{\institution{Akamai/Hong Kong PolyU}
  \city{Hong Kong}
  \country{China}
}

\author{Roland Dobbins} 
\affiliation{\institution{NETSCOUT}
  \city{Westford}
  \state{MA}
  \country{USA}
}

\author{Christian Doerr} 
\affiliation{\institution{Hasso Plattner Institute}
  \city{Potsdam}
  \country{Germany}
}

\author{Christian Rossow} 
\affiliation{\institution{CISPA}
  \city{Saarbr\"ucken}
  \country{Germany}
}

\author{Daniel R. Thomas} 
\affiliation{\institution{University of Strathclyde}
\city{Glasgow}
\country{Scotland}
}

\author{Mattijs Jonker} 
\affiliation{\institution{University of Twente}
  \city{Enschede}
  \country{The Netherlands}
}

\author{Ricky Mok} 
\affiliation{\institution{CAIDA/UC San Diego}
  \city{La Jolla}
  \state{CA}
  \country{USA}
}

\author{Xiapu Luo} 
\affiliation{\institution{Hong Kong PolyU}
  \city{Hong Kong}
  \country{China}
}

\author{John Kristoff} 
\affiliation{\institution{NETSCOUT/UIC}
  \city{Westford}
  \state{MA}
  \country{USA}
}

\author{Thomas C. Schmidt} 
\affiliation{\institution{HAW Hamburg}
  \city{Hamburg}
  \country{Germany}
}

\author{Matthias W\"ahlisch} 
\affiliation{\institution{TU Dresden}
  \city{Dresden}
  \country{Germany}
}

\author{kc claffy} 
\affiliation{\institution{CAIDA/UC San Diego}
  \city{La Jolla}
  \state{CA}
  \country{USA}
}

\begin{abstract}

Motivated by the impressive but diffuse scope of DDoS research and reporting,
we undertake a multistakeholder (joint industry-academic)
analysis to seek convergence across the best available
macroscopic views of the relative trends in two dominant classes
of attacks – direct-path attacks and reflection-amplification attacks.
We first analyze 24 industry reports to extract trends and (in)consistencies
across observations by commercial stakeholders in 2022.
	We then analyze ten data sets spanning industry and academic sources,
across four years (2019-2023), to find and explain discrepancies based on data
sources, vantage points, methods, and parameters.  
Our method includes a new approach:  we share an aggregated list of
DDoS targets with industry players who return the results of joining
this list with their proprietary data sources to reveal gaps in visibility
of the academic data sources.  We use academic data sources to explore
an industry-reported relative drop in spoofed reflection-amplification
attacks in 2021-2022.  Our study illustrates the value, but also
the challenge, in independent validation of security-related 
properties of Internet infrastructure. 
Finally, we reflect on opportunities to facilitate greater
common understanding of the DDoS landscape.
We hope our results inform not only future
academic and industry pursuits but also emerging policy efforts
to reduce systemic Internet security vulnerabilities.

\end{abstract}

\maketitle
\renewcommand{\shortauthors}{Raphael Hiesgen et al.}

 \definecolor{boxgray}{rgb}{0.93,0.93,0.93}
 \textblockcolor{boxgray}
 \setlength{\TPboxrulesize}{0.7pt}
 \setlength{\TPHorizModule}{\paperwidth}
 \setlength{\TPVertModule}{\paperheight}
 \TPMargin{5pt}
 \begin{textblock}{0.8}(0.1,0.04)
   \noindent
   \footnotesize
   If you refer to this paper, please cite the peer-reviewed publication: Raphael Hiesgen, Marcin Nawrocki, Marinho
   Barcellos, Daniel Kopp, Oliver Hohlfeld, Echo Chan, Roland Dobbins, Christian Doerr, Christian Rossow, Daniel R.
   Thomas, Mattijs Jonker, Ricky Mok, Xiapu Luo, John Kristoff, Thomas C. Schmidt, Matthias Wählisch, KC Claffy.
   2024. The Age of DDoScovery: An Empirical Comparison of Industry and Academic DDoS Assessments. 
   In \emph{Proceedings of ACM Internet Measurement Conference (IMC)}. ACM, New York. https://doi.org/10.1145/3646547.3688451
\end{textblock}

\section{Introduction}

Distributed Denial-of-Service (DDoS) attacks were first reported around
2000~\cite{paxson2001reflectors,changDefendingFloodingbasedDistributed2002}
and continue to cause substantial damage, with cycles of new
attack strategies and novel mitigation approaches.  
While hundreds of scientific studies and proposals have provided academic perspectives (\eg~\cite{r-ahrnp-14,rowrs-adads-15,griffioen2021adversarial,nawrocki2021ecosystem,nkkhs-advmd-23}),
the more impactful developments have been commercial, where the need to mitigate the harms of DDoS 
led to the creation of a vibrant DDoS mitigation market
worth over US\$~1.5
Billion~\cite{futuremarketinsightsfmiDDoSProtectionMarket2023} and multiple practical attempts to deter a broad range of
attacks~\cite{RFC-2827,luckieNetworkHygieneIncentives2019,europolGlobalCrackdownDDoS2022,shadowserverDDoSShadowserverFoundation2023}.
The concentration of content services among a 
few heavily provisioned network infrastructures 
has also provided some protection against the threat of DDoS, at the cost of strengthening their oligopolies.
This combination of industry forces that benefit from DoS
prevalence has arguably reduced the motivation for collective action
to remediate the underlying DDoS threat and its root causes.

While academic projects have attempted longitudinal analysis of
DDoS trends, gaining a consensus view of the state of the DDoS 
landscape has proven elusive.
We undertake an extended multi-stakeholder 
analysis to pursue such consensus. 
We focus on direct-path attacks and
reflection-amplification attacks, two dominant classes of attacks. 
We define each attack class and approaches to detecting
and mitigating them (\autoref{sec:taxonomy}). 
We analyze the state of industry reporting on this topic, reviewing 
24 reports to extract trends and 
inconsistencies across them (\autoref{sec:industry-reports}).  
We then describe the range of (raw and derivative) data sources
available and challenges in comparing them, confirming that
different detection approaches, and even the same approach
using different parameters and vantage points, will yield
different inferences of attack scope, duration, and
impact (\autoref{sec:comparing-data-sources}).

We analyze ten data sets covering a four years' period (2019--2023) to explore discrepancies
based on data sources, vantage points, methods, and parameters
(\autoref{sec:long-term-trends}, \autoref{sec:targets}).
In the process, we use academic data sources to explore 
an industry-reported relative drop in spoofed reflection-amplification 
attacks following a concerted industry effort to encourage 
deployment of source address validation (SAV).
However, we find more differences than similarities across data sets.
\autoref{tbl:overview-results} summarizes partial inconsistencies visible across various DDoS observatories used in this paper, and also among industry reports (from $\approx$ 2022). 
Our work reinforces findings of \cite{nkkhs-advmd-23} that  singular data sources may have serious visibility
	limitations, which provide the strongest empirical grounding to date for regulatory framing to share data.
Our four contributions are:
\begin{enumerate}
\item We taxonomize information extracted from industry 
reports characterizing DDoS phenomena in 2022-2023, which 
we publish as supplementary knowledge base, including
an archive of the reports (\S\ref{sec:industry-reports}).  
\item We quantitatively compare ten data sources over four years, 
spanning honeypots, IXPs, and edge networks, including
industry and academic vantage points.  To
our knowledge this is the largest correlation of longitudinal DDoS data
ever published~(\S\ref{sec:data},\ref{sec:long-term-trends}).
\item  We propose and execute a new approach to facilitating a degree
of industry transparency, by aggregating academic sources and 
sharing them with industry players who then return the results of
joining these shared data sets with their proprietary data sources
to indicate gaps in visibility of the academic data 
 (\S\ref{sec:targets}).
\item We propose several recommendations 
to facilitate scientific study of the DDoS landscape
and of whether proposed mitigations are effective. We introduce possible self-regulatory approaches, other potential regulatory developments, and roles for
 researchers (\S\ref{sec:future-ddos}).
\end{enumerate}

\begin{table*}
  \center
  \caption{Data comparison results: Partially inconsistent views among DDoS data observatories used in this paper measuring decreasing~\downwardTrend\ ($< -5\%$ in 4 years), increasing~\upwardTrend\ ($>5\%$ in 4 years), and steady~\neutralTrend\ trends of attack types in 2019--2023. The surveyed industry reports from $\approx$ 2022, which usually compare relative share of attacks, similarly provide inconsistent views. Here, numbers in braces indicate the number of reports out of 24~surveyed reports.}
  \label{tbl:overview-results}
  \begin{tabular}{lccccccccc}
  \toprule

  Attack Type & \multicolumn{8}{c}{Observatories Used in This Paper (2019-2023)} & Industry Reports (\#)
  \\
  \cmidrule(lr){2-9}
  & \multicolumn{2}{c}{Network Telescopes} & \multicolumn{3}{c}{Flow Data} & \multicolumn{3}{c}{Honeypots} & ($\approx$ 2022)
  \\
  \cmidrule(lr){2-3}
  \cmidrule(lr){4-6}
  \cmidrule(lr){7-9}
  & UCSD & Orion & Netscout & Akamai & IXP & Hopscotch & AmpPot & NewKid &
  \\
  \midrule
  Direct-path & \upwardTrend & \upwardTrend & \upwardTrend & \neutralTrend & \upwardTrend & n/a & n/a & n/a & \upwardTrend (5), \downwardTrend(0)
  \\
  Reflection-Ampl. & n/a & n/a  & \upwardTrend & \neutralTrend & \downwardTrend & \downwardTrend & \neutralTrend & \upwardTrend & \upwardTrend (2), \downwardTrend (3)
  \\
  \bottomrule
  \end{tabular}
\end{table*}

Our empirical study clearly shows that the assessment of DDoS trends is challenging and that collaboration between research and industry is needed to gain sound insights.
Because regulators are now showing vigorous interest in policy intervention
to reduce systemic Internet security vulnerabilities 
\cite{nis2-eu-2021,EU-cyber-resilience-act-2023}, we hope that the 
results of this paper help inform and guide
not only academic and industry efforts but also future policy decisions.

\section{Definitions, Detection, Defense}
\label{sec:taxonomy}

We describe the two most prevalent classes of DDoS attacks---\one direct-path (spoofed and non-spoofed) attacks and \two reflection-amplification attacks---and 
methodologies to observe them.
For a broader classification of DDoS~attacks, we refer to prior work
\cite{mr-tdadd-04, dm-dadmc-04, sl-ddsta-04, m-dmadd-18}.

\subsection{DDoS Attack Models}
\label{subsec:ddos-models}

\begin{figure}
  \center
  \includegraphics[width=1\columnwidth]{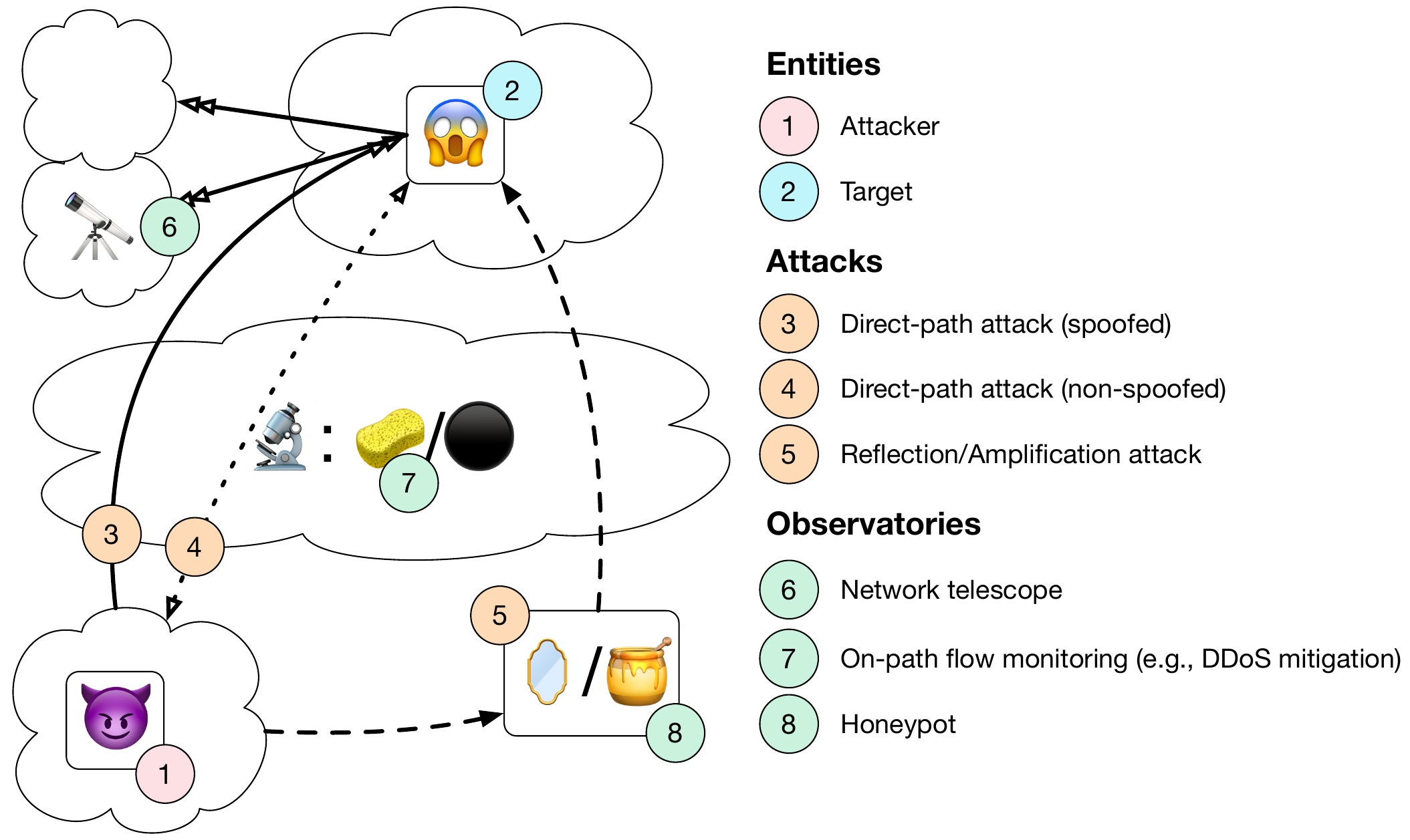}
	\caption{Three DDoS attack types: Direct-path spoofed (solid line), direct-path non-spoofed (dotted line), and typically spoofed reflection-amplification (dashed).}
  \label{fig:overview}
\end{figure}

\autoref{fig:overview} illustrates direct-path and reflection-amplification
attacks.  In either cases, an attacker \circled{1} aims to overwhelm a
target host or target service \circled{2}, or saturate its uplink.

\paragraph{Direct-path attacks} Attackers send packets
directly to the target. If the source address is
\textit{spoofed}~\cite{b-sptps-89} \circled{3}, the target sends responses
 to the hosts with the spoofed addresses.
An example is the well-known {\em SYN-flood attack}~\cite{RFC-4987}, where an
attacker sends TCP SYN packets, each of which induces a memory allocation
related to the TCP connection.
The spoofed
source addresses are often chosen randomly, leading to the term {\em
randomly-spoofed DDoS (RSDoS) attacks}.

Another type of direct-path attack uses \textit{non-spoofed} source
addresses \circled{4} to establish many sustained connections with
a server~\cite{ddlml-hdsle-12}. 
This state exhaustion attack also minimizes impact on the
sending network and visibility of the attack. 

\paragraph{Reflection-amplification attacks} The attacker indirectly
sends traffic to a target via a \textit{reflector} \circled{5}.
The reflector typically produces large responses
to small requests, \ie it \textit{amplifies}~\cite{r-ahrnp-14}. Amplifiers
allow attackers to subject their target to massive amounts of traffic 
by sourcing much less traffic from their own network,
reducing the impact as well as likelihood of detection of the~attack. 

\paragraph{Enabling platforms}
Attackers today frequently rent facilities on
\textit{botnets} or dedicated infrastructure, both of which are robust
to takedowns and hide the attacker's identity.  \textit{Bullet-proof
hosters} (BPH) are service providers that avoid responding
to law enforcement requests and are lenient with acceptable use.
\textit{Booter} services perform DDoS attacks for a fee.
They are surprisingly resilient to takedowns or after takedown often return shortly
on a new website~\cite{BooterShutdown,kopp2019hide}.

\subsection{DDoS Observatories}
\label{sub:ddos-observing}

We describe three types of measurement that allow inferences about DDoS attacks:
network telescopes,
flow monitoring,
and honeypots.

\paragraphNoDot{\telescope{} Network telescopes (NT)}~\circled{6} are passive measurement platforms that collect packets sent to
large blocks of unused IP space. Network telescopes
achieve visibility of attack preparation in the form
of scans for open reflectors or vulnerable hosts.  Telescopes also collect
artifacts of certain types of attack execution, \ie replies
to randomly spoofed packets in RSDoS attacks. By identifying typical
response packets and thresholds, network telescopes enable inference of
scope, prevalence, and duration of RSDoS attacks~\cite{2006-moore-bd, jkkrs-mtuam-17}.
Network telescopes do not generally observe evidence of 
reflection-amplification attacks, since the spoofed address in the
attack traffic is not random, but rather the address of 
the intended target. The attack traffic is also destined to the
reflector, rather than broadly toward the Internet where 
a telescope might observe it. 

\paragraph{\dpi{} On-path flow monitoring by DDoS mitigation providers}
Observing DDoS traffic toward
victims requires a vantage point \circled{7} on the path to the target. DDoS identification can
rely on manual inference, deep-packet inspection, aggregate packet- or flow-level
statistics~\cite{kdh-dndip-21}, or more complex 
machine learning approaches~\cite{ixpscrubber}.
IXPs, CDNs, or specialized DDoS protection service (DPS) providers
may protect their customers 
by {\em scrubbing}~\scrubbing{} or {\em blackholing}~\blackhole{} attack traffic (\autoref{subsec:history}).
These vantage points directly
observe ongoing attacks, but access to such data is limited to the network
owner or commercial DDoS mitigators serving that~network.

\paragraphNoDot{\honeypot{} Honeypots (HP)}~\circled{8} emulate a vulnerable host 
to learn more about the behavior of an attacker~\cite{ph-vhfbt-08,
nkkhs-advmd-23}. 
The level at which a honeypot interacts with a presumed attacker
ranges from one packet to full compromise and access to a
service. To support the study of DDoS, honeypots may try to appear
as reflectors for common protocol vectors, \eg DNS
or NTP.  To avoid participating in attacks, honeypots
stop engaging with a probing source after some sending threshold is reached.  Several honeypot platforms
have been operational for years (\S\ref{sec:long-term-trends}), \eg
AmpPot~\cite{kramer2015amppot}, AmpPotMod~\cite{noroozian2016amppot},
Hopscotch~\cite{thomas2017thousand}, NewKid~\cite{heinrich2021kids},
and HPI~\cite{griffioen2021adversarial}.

\subsection{Prevention and Mitigation of DDoS}
\label{subsec:history}

We review approaches to prevention and mitigation 
of the two DDoS types we study: disabling amplifiers, anti-spoofing
campaigns, booter takedowns, and filtering of attacks.

\paragraph{Prevention: Disabling reflection-amplification vectors}
Academic and industry efforts to identify
and decommission open servers that support reflection and
amplification\cite{nawrocki2021transparent,kuhrer2014exit,
ShadowserverReporting,netscout-TP240-2022,netscout-SIP-RA-2023}
have had limited success.  Some services
are easier to decommission than others, \eg operators of NTP servers
can disable a specific command that enables extraordinary amplification
({\tt get monlist}) but is not of operational use.
DNS servers are less amenable to such curation of function.
DNS operators must consider more complex configuration changes, 
such as rate limiting, filtering (e.g., \texttt{ANY} requests), 
or truncating large responses. 
The result is a long-standing persistence of amplification vectors
\cite{kdh-dndip-21,nawrocki2021transparent}.\footnote{Authors 
of \cite{nawrocki2021transparent} continue measuring the prevalence of
transparent DNS forwarders, showing a drop in mid-2023.  \url{https://odns.secnow.net/data} 
}

\paragraph{Prevention: Promotion of source address validation (SAV)}
Operators have pursued efforts to reduce the
number of networks that allow source IP address spoofing, the basis
of all spoofed DDoS attacks.  One such effort is the 
Spoofer measurement project~\cite{luckieNetworkHygieneIncentives2019}, which identifies networks that allow spoofing and assists with
remediation of this vulnerability. This project relies on users 
to download software from CAIDA's website and launch it in
the background on their laptop; the software tests each
new network visited for the ability to spoof.  This volunteer
crowdsourced approach yields limited measurement coverage.  

For many years
groups have undertaken various efforts to eliminate
sources of spoofing in networks~\cite{dtfan-mymmm-22}. 
DDoS mitigation providers reported a successful concerted effort 
since 2021 by the global Internet operational community to 
reduce spoofing~\cite{arelionArelionDDoSThreat2023, nokiaChangingDDoSThreat2022,netscoutNetscoutDDoSThreat2023a}.\footnote{``..the lower global backbone impact was largely due to an industry wide antispoofing initiative -- the DDoS Traceback Working Group.'' \cite{nokiaChangingDDoSThreat2022}, s.a.~\cite{m3aawg-ddosaward}.}
Netscout reported a 17\% decrease in reflection-amplification 
attacks (which leverage spoofable networks) in 2022 
compared with 2021, in their view a direct result of 
this concerted effort \cite{netscout2023April-archive}.\footnote{``In 2022 [...] a momentous 17 percent global decrease in reflection/amplification attacks was observed when compared with 2021. '' \cite{netscout2023April-archive}.}
Our observatories also saw drops between 2021 and 2022, 
of varying magnitudes (see \autoref{fig:normalized_attack_trends:reflampl}
and discussion in \autoref{sub:analysis:reflampl}).

\paragraph{Prevention: Takedown of booter services by law enforcement}
Other efforts by academia, industry, and law enforcement
have focused on shutting down DDoS-for-hire 
services~\cite{kopp2019hide,BooterShutdown,krebs-may2023booters,dismantling-ddos-blackhat-2023}.
Booters often reappear within a few months under different domains,
seemingly leaving little long-term reduction from the effort~\cite{BooterShutdown}.
Our analysis of observed reflection-amplification attacks 
(\autoref{sub:analysis:reflampl}) shows no
lasting downward trend subsequent to recent publicized 
large-scale takedowns~\cite{krebs-may2023booters,Takedown202305,usdc-swtor-23}.
Law enforcement have also used targeted messaging campaigns (Google search result ads for keywords like `booter') to raise awareness that these websites offer illegal services. 
This approach may be a cheap and effective way of reducing DDoS~\cite{BooterShutdown}, but its effectiveness remains unclear~\cite{moneva_effect_2023},
as do the ethics of this approach~\cite{collier_influence_2022}.

\paragraph{Mitigation: Filtering attack traffic}
Filtering ongoing attacks at the victim's network 
allows for a close loop between detection and mitigation
but limits the scope of mitigation since the victim network must
still receive attack traffic.
A more effective approach is a {\em scrubbing service}, where a third
party (\eg IXP, CDN, or DPS) uses deep-packet inspection or application
proxies to identify and block DDoS attacks at network/application layers
and forward the sanitized traffic to their customer's networks
~\cite{dietzelStellarNetworkAttack2018,ixpscrubber,tung18bgpdps,sbkms-adpaa-20,jonker2016protection}.
A coarser-grained approach is
remote triggered black hole (RTBH) filtering~\cite{rfc5635,gsdrf-ibbai-17,nawrocki2019down,kdh-dndip-21,jonker2018-blackholing},
where a target (victim) remotely triggers the dropping of traffic 
to a whole IP prefix when one or more addresses in that prefix
is under a DDoS attack.  Blackholing risks
collateral damage~\cite{nawrocki2019down,jonker2018-blackholing}.

\paragraph{Mitigation: Standardization efforts to support cooperative filtering}
Standards for DDoS defense and prevention
are partially documented as IETF best current practice (BCP)~\cite{BCP84RFC3704,BCP84RFC8704}, informational RFCs
\cite{rfc3954,rfc5635}, and standards-track RFCs~\cite{rfc8955,rfc7011}.
Other proposed standards with weaker operational roots did not gain
traction \cite{rfc2748,rfc8612,rfc9387,x805_2003}.
Operator groups have published their own BCPs \cite{m3aawg-recommendations-2017}, and two
ISPs presented a bilateral {\em DDoS peering} framework
to allow mutual filtering of DDoS traffic between peers
\cite{levy-nanog-2017,compton-nanog-2019}.
In 2020, DE-CIX proposed a technical and governance
framework to facilitate DDoS-related data
sharing among Internet Exchange Points (IXPs) \cite{DXP}.
The Netherlands Anti-DDoS Coalition (ADC) has pursued a
similar concept at national scale~\cite{dutchAntiDDoS}.
Team Cymru has created a service to 
facilitate relaying of destination-based remote triggered
black holing (RTBH) signals between ASes~\cite{k-caris-15,UTRS}.
Deployment, scaling, and sustaining such collaborative efforts have 
proven challenging~\cite{Anghel2023Peering}.

\section{Analyzing DDoS Industry Reports}
\label{sec:industry-reports}

The DDoS mitigation industry publishes reports about the state of DDoS, identifying trends and alerting decision-makers about the need to deploy appropriate DDoS protection.
The primary purpose of these reports is to promote use of a DDoS mitigation approach.
They also offer a glance at data usually not accessible to researchers. %
Since industry and academia have diverging views~\cite{nkkhs-advmd-23}, we were motivated to take a close look at published threat reports.
We dissected 24~reports of 22~vendors to contrast numbers and trends, laying a foundation to compare industry with academic perspectives.
We undertook and now publish a thorough \textit{structural analysis of these reports as a supplementary artifact} \cite{ddos-industry-github}; due to space constraints, we only summarize highlights here.

\paragraph{Our method to survey industry reports on DDoS}
DDoS reporting from industry is fragmented, scattered among periodic reports, blogs, related educational resources, and talks. 
We limit our analysis to written content, which we call ``reports''.
We collected reports from companies listed in a related market survey~\cite{rm-gdpms-23} and excluded global threat reports without DDoS content~(\eg \cite{fortinetGlobalThreatLandscape2023,crowdstrike2023GlobalThreat2023,paloaltoUnit42INCIDENT}), DDoS reports without DDoS data~(\eg \cite{fastlyWhatDDoSAttack2023,rioreyRioReyTaxonomyDDoS2015}), and DDoS assessments before 2022~(\eg \cite{awsShieldThreat2021,nokiaNokiaDeepfieldNetwork2022}).
We considered available reports from all major DDoS mitigation providers: A10~\cite{a102022A10Networks2022}, Akamai~\cite{akamaiRelentlessEvolutionDDoS2022}, Arelion~\cite{arelionArelionDDoSThreat2023}, Cloudflare~\cite{cloudflareDDoSTrends2023}, Comcast~\cite{comcast2023ComcastBusiness2023}, Corero~\cite{corero2023DDoSThreat2023}, DDoS-Guard~\cite{ddos-guardDDoSAttackTrends2023,ddos-guardDDoSGuardAnalyticalReport2023}, F5~\cite{f5F5DDoSAttack2023}, Huawei~\cite{huaweiGlobalDDoSAttack}, Imperva~\cite{impervaImpervaGlobalDDoS2023}, Kaspersky~\cite{kasperskyKaperskyDDoSAttacks2022}, Link11~\cite{link11LINK11DDOSREPORT20222023}, Lumen~\cite{lumenLumenQuarterlyDDoSQ42022}, Microsoft Azure~\cite{microsoftazurenetworksecurityteam2022ReviewDDoS2023}, NBIP~\cite{nbipDDoSAttackFigures2023}, Net\-scout~\cite{netscout5thAnniversaryDDoS2023}, NexusGuard~\cite{nexusguardDDoSStatisticalReport2023}, Nokia~\cite{nokiaNokiaThreatIntelligence2023}, NSFocus~\cite{nsfocus2022GlobalDDoS2023}, Qrator~\cite{qratorQ42022DDoS23}, Radware~\cite{radwareRadwareGlobalThreat2023}, and Zayo~\cite{zayoProtectingYourBusiness2023}.
Most reports were released early 2023 and focus on 2022. 
\autoref{appendix:industry-reports} provides details.

\paragraph{Presentation style}
Industry reports are unlike scientific papers, typically using vague language and lacking clarity about data analysis methodologies. 
They vary substantially in format and organization, from full documents to web blogs to infographics.
Some reports cover DDoS exclusively (\eg \cite{corero2023DDoSThreat2023,cloudflareDDoSAttackTrendsQ42022,f5F5DDoSAttack2023,huaweiGlobalDDoSAttack,netscout5thAnniversaryDDoS2023}), while others report on a range of malicious activities (\eg \cite{comcast2023ComcastBusiness2023,nokiaNokiaThreatIntelligence2023,qratorQ42022DDoS23}).
Technical depth spans from superficial trends to in-depth analyses that explain the vectors and methods to launch attacks.
Not even the most detailed reports clearly explain the methodologies to identify attacks or discuss limitations of their analyses.
Industry reports also do not contextualize the findings in terms of overall traffic patterns so that readers cannot judge whether attacks are growing in proportion with other properties, \eg user base. 
Reports mix absolute and relative values, depending on the message to be emphasized. 
For example, a high increase (\eg 500\%) in some form of attack may actually represent a small absolute 
change~\cite{netscoutNETSCOUTDDoSAttack2023}. 

What concerns us most is that some reports cherry-pick numbers to convince readers about the increase of DDoS attacks and the damage they cause.
Most reports highlight the growing impact of DDoS attacks, but when the data suggests a decrease in severity, the message is less clear. 
Marketing concerns may lead to revision of reports prepared by technical staff, to present observations in a way that is more aligned with business interests.

\paragraph{Metrics used by reports}
The reports we analyzed used a range of numbers to illustrate the DDoS attack landscape (in 2022) compared to previous periods. 
The attack attributes frequently reported by industry papers are:
  \emph{count}: per period and attack type;
  \emph{size}: peak packet rate, peak bandwidth, or attack volume; %
  \emph{duration}: in minutes/hours (\eg ``most attacks under 10 min''); %
  \emph{vectors}: protocol/packets used (\eg TCP SYN and DNS amplification);
  \emph{methods}: carpet-bombing~\cite{nexusguardDDoSStatisticalReport2023}, 
  	 pulse-wave;
  \emph{vector instances}: number of hosts that can send attack packets;
	and \emph{context}, \eg cyberwarfare or hacktivism.
Some reports include information about the use of multi-vector attacks, attack repetition, use of botnets, targeted industries (\eg finance sector, IT, education), and geolocation of attack sources and targets.

\paragraph{Analysis period}  
Most reports focus on one year, comparing with the previous year or sometimes a few years back if highlighting a trend.
Generally, metrics show oscillations when the period used in a report is a quarter or a month. 
Comparing short periods may be misleading, but may illustrate cyclic behaviors, \eg use of a vector every few months.

\paragraph{Comparing findings}
Companies generally reported an overall increase in DDoS attacks. 
Among the exceptions, F5 indicated a decrease of 9.7\% in total attacks~\cite{f5F5DDoSAttack2023}, while Arelion reported a ``dramatic'' reduction of DDoS activity~\cite{arelionArelionDDoSThreat2023}.
Arelion associated the drop with a decrease in UDP spoofed attacks, following actions of an ``industry-wide anti-spoofing initiative'' \cite{arelionArelionDDoSThreat2023} (\autoref{subsec:history}), and despite some increase in direct path attacks.  
Netscout also reported a drop in reflection-amplification attacks, and Akamai reported a decrease in CharGEN, SSDP and CLDAP-based attacks, typical amplification vectors. 
Several providers, namely Cloudflare, F5, Imperva, NBIP, Netscout, NexusGuard, and Radware, reported substantial increases in application-layer (L7) attacks, \eg via HTTP/S.
Consistency across most reports is the dominance of UDP-based vectors, predominantly UDP~flooding.

\paragraph{Summary}
The reports we reviewed reflect a variety of data sources and analysis methods.
Industry reports can provide directions for researchers to perform more scientific explorations, but they do not present or claim a scientific contribution.
Given pecuniary interests and perspectives of vendors that publish reports, our community should consider them with care. 
Reported results can complement scientific studies but still do not provide a complete picture.
Our analysis of these reports (see \autoref{appendix:industry-reports} and \cite{ddos-industry-github}) further inspired our development of a more rigorous framework for comparison. 

\section{Challenges in Comparing DDoS Data}
\label{sec:comparing-data-sources}

Finding consensus on the DDoS landscape faces three obstacles: 
vantage point limitations; definitions and detection methods;
and inhibitions on data sharing.  

\paragraph{Vantage point limitations}
Characteristics of observed DDoS attacks vary by observation point
and method (\S\ref{sub:ddos-observing}).
DDoS inference in the Internet core, \eg from flow data
at IXPs, more likely 
underestimates attack length and volume of observed attacks,
since some (or all) attack traffic may transit paths other than the IXP. 
In contrast, a vantage point at a DDoS protection service (DPS) 
or the Internet \text{edge}, \eg
a victim's network,  observes only packets targeting or originating from the 
observed network(s). 
Measuring close to the target 
can lead to more accurate detection, but 
obstacles to sharing such data.  
 Concerns regarding privacy or reputation may  
prevent its use for establishing a consensus view. 
Researchers have developed aggregated vantage 
points (honeypots and telescopes)
that have lower obstacles to sharing data.
Honeypots observe reflection-amplification attacks when
an attacker selects the honeypot as a reflector for the attack.  
Telescopes observe RSDoS backscatter, and if
monitoring a large enough segment of address space,
they achieve high visibility of such
attacks independent of target~location.

\paragraph{Definition of attack}
Distinguishing between natural traffic peaks and attack traffic is challenging.
Honeypots need to discern
scanning~\cite{kramer2015amppot} and testing by attackers from actual
attacks.  No single definition can accurately capture
characteristics of all attacks.  Appropriate thresholds
depend on attack type, protocol~\cite{nkkhs-advmd-23}, and
observatory~\cite{kdh-dndip-21}.  
Some attacks require additional inference and sanitization steps,
including aggregation across multiple sensors~(\S\ref{sec:data}).

\paragraph{Data sharing obstacles}
Prohibitions on multi-lateral (vs bi-lateral) data sharing 
create obstacles to mitigation. 
Personal trust between individual members
does not automatically translate to trust in procedures
and governance of a larger collaborating 
group~\cite{gommans2015service,geras-2023}. 
The key challenge is establishing a data governance model
and legal agreements to support it. 
Some established such agreements successfully~\cite{vdhout2022ddosch,CCCCDataAgreements}.

\paragraph{Longitudinal trend bias} 
We used normalized attack counts per week (\autoref{sec:long-term-trends}),
without considering growth in traffic, customers, or measurement coverage.
Normalization reduces data sharing concerns and helps control for
coverage variation among data sources.

\begin{table*}[t]
\setlength{\tabcolsep}{4.5pt}
  \caption{The observatories used in this research vary in collection methods and attack detection strategies. Honeypots use different flow identifiers, see~\cite{nkkhs-advmd-23}. (Location: \textbf{G}eographically \& \textbf{T}opologically distribution.)}
  \label{tab:vantagepoints}
  \centering
  \begin{tabular}{l c c c c l l l}
    \toprule
    Platform & Type & Attack & Loc. & Coverage & \multicolumn{3}{c}{Attack Definition} \\
    \cmidrule(lr){6-8}
                  &      &        &          &          & Flow Identifier & Timeout & Threshold \\
    \midrule
    UCSD NT                        & \inlinegraphics{emojis/telescope.png} & RSDoS & US         & ~12M IPs   & protocol, src IP & 300s & $\geq$ 25 pkts, $\geq$ 60s$^2$ \\
    ORION NT                       & \inlinegraphics{emojis/telescope.png} & RSDoS & US         & 500k IPs   & protocol, src IP & 300s & $\geq$ 25 pkts, $\geq$ 60s$^2$ \\
    \midrule
    Netscout Atlas (RA)                   & \inlinegraphics{emojis/dpi.png} & DP    & G/T &  proprietary           & \multicolumn{3}{l}{Hand-craft flow identifiers \& thresholds} \\
    Netscout Atlas (DP)                   & \inlinegraphics{emojis/dpi.png} & RA    & G/T &   proprietary          & \multicolumn{3}{l}{Hand-craft flow identifiers \& thresholds} \\
    Akamai Prolexic (RA) & \inlinegraphics{emojis/dpi.png} & RA    & G/T  &  proprietary           & \multicolumn{3}{l}{Hand-craft flow identifiers \& thresholds} \\
    Akamai Prolexic (DP) & \inlinegraphics{emojis/dpi.png} & DP    & G/T  &  proprietary           & \multicolumn{3}{l}{Hand-craft flow identifiers \& thresholds} \\
    IXP BH (RA)\cite{kdh-dndip-21} & \inlinegraphics{emojis/dpi.png} & DP    & G/T  &  proprietary           & UDP, ampl. src port & & $\geq10$ IPs, $>1$ Gbps \\
    IXP BH (DP)\cite{kdh-dndip-21} & \inlinegraphics{emojis/dpi.png} & RA    & G/T  &   proprietary          & TCP & &  $\geq10$ IPs, $>100$ Mbps \\
    \midrule
    AmpPot~\cite{kramer2015amppot} & \inlinegraphics{emojis/honeypot.png} & RA    & G/T & $\approx$ 30 IPs & Src IP, src port, dst IP, dst port               & 60 min & $\geq100$ pkts \\
    Hopscotch~\cite{thomas2017thousand}  & \inlinegraphics{emojis/honeypot.png} & RA    & G/T & 65 IPs    & Src IP, dst IP, dst port       & 15 min & $\geq5$ pkts \\
    NewKid~\cite{heinrich2021kids} & \inlinegraphics{emojis/honeypot.png} & RA    & BR         &  1 IP     & Src prefix, dst IP, [dst port]$^1$ & 1 min  & $\geq5$ pkts, [$\geq2$ ports]$^1$ \\
	\bottomrule
	\multicolumn{8}{l}{\footnotesize $^1$ NewKid uses two thresholds, one for mono-(dst port) and for multi-protocol ($\geq2$ ports) attacks. $^2$ See \autoref{sec:app-rsdos-infer} for RSDoS inference details.} \\
  \end{tabular}
\end{table*}

\section{Data Corpus Used in This Study}
\label{sec:data}

We analyze ten data sets from seven observatories: 2 network telescopes (\inlinegraphics{emojis/telescope.png}), 3 DDoS mitigation providers (\inlinegraphics{emojis/dpi.png}), and 2 honeypots (\inlinegraphics{emojis/honeypot.png}).
\autoref{tab:vantagepoints} summarizes information about these observatories such as the types of attack they measure: direct path (DP), reflection-amplification (RA), or randomly-spoofed DoS (RSDoS), a subset of DP attacks (\autoref{sub:ddos-observing}).
We compare attack data across 4.5 years (2019 to mid-2023).
To our knowledge this is the largest correlation of DDoS data sets not only including academic but also industry sources.
Ethical considerations are discussed in \autoref{appendix:ethics}.

Each observatory captures different traffic, and thus may not
see the same attack at the same intensity, or
at all.  This distinction is often a function of the vantage point
(\autoref{sec:comparing-data-sources}).  Honeypots and telescopes
are essentially end points in some portion of attack traffic,
whereas industry (traffic flow-based) solutions sit somewhere on the
network path, perhaps toward the target endpoint. Flow data 
will include not just attack-induced traffic but legitimate traffic which
can offer a baseline for comparison and inference. 

Another concern is that observatories might interfere with each other's
visibility. For example, an observed but quickly mitigated randomly-spoofed
direct-path attack might not reflect packets into a network telescope.
Partially mitigated attacks may affect the attack proportions inferred
by other observatories (\eg length, volume).

Finally, the attack detection strategy, including threshold parameters, defines
how an observatory identifies DDoS attack in its traffic.  Lack of ground
truth data on attacks prevents confident estimates of DDoS detection
accuracy.  Academic sensors may over-estimate attacks, \eg if honeypots
mistakenly interpret scans as attacks.  In contrast, industry efforts
to mitigate large attacks may miss short or low-volume attacks that
nonetheless harm a target, \eg an uplink of an end user.

\paragraph{\inlinegraphics{emojis/telescope.png} Telescopes observe (backscatter from randomly-sp\-oof\-ed) direct path attacks}
We used data from two of the largest, longest-running IPv4 network telescopes: Merit's \textit{ORION project}~\cite{orionWebsiteTelescope} with $\approx$500k IPv4 addresses and UCSD's \textit{UCSD-NT} operated by CAIDA \cite{caidaWebsiteTelescope}, spanning a lightly utilized /9 and /10 network, \ie $\approx$12M IPv4 addresses. Telescopes (\circled{6} in \autoref{fig:overview}) observe backscatter from direct path attacks that use randomly-spoofed source addresses. %
Assuming an approximately random distribution of spoofed IP addresses,
larger telescopes will receive more attack traffic and can thus
detect smaller attacks.  Using the parameters suggested in \cite{2006-moore-bd},
\textit{UCSD-NT} and \textit{ORION} can detect DDoS events with attack
rates of 0.026 Mbps and 0.60 Mbps in 5 minutes, respectively. 
(With the same assumptions, a
/20 telescope could detect attacks of $\sim$70Mbps in 5 minutes.)

For both telescopes we had the raw traffic data available.
We used the RSDoS-detection algorithm developed at CAIDA (based on \cite{2006-moore-bd}) on both data sets to identify attacks, which were the basis for our analysis.  \autoref{sec:app-rsdos-infer} has details of CAIDA's current algorithm.

\paragraph{\inlinegraphics{emojis/dpi.png} Monitored flow data can include both attack types}
We had access to DDoS attack counts from two DDoS mitigation providers (\circled{7} in \autoref{fig:overview}).
Both observe traffic in an on-path network. The first data set, \textit{IXP Blackholing}, contains daily counts of attacks identified for traffic that was blackholed by a European IXP (method in~\cite{kdh-dndip-21}). 
The second data set contains daily attack counts observed by Netscout, which receives anonymized DDoS attack statistics from more than 500 ISPs and 1500 enterprises worldwide.
These industry sources do not share data that would reveal anything about 
a customer suffering an attack.
We received the daily attack counts separated by attack type (reflection-amplification and direct-path).
Netscout also provided counts for spoofed and non-spoofed attacks in the DP attack~data.
The third data set was collected by Akamai Prolexic, a DDoS protection service (DPS) that detects and mitigates attacks in traffic transiting its AS.
It includes weekly attack counts.

\begin{figure}
  \begin{subfigure}{.5\textwidth}
    \centering
    \includegraphics[trim={0 0.2cm 0 0},clip,width=1.0\linewidth]{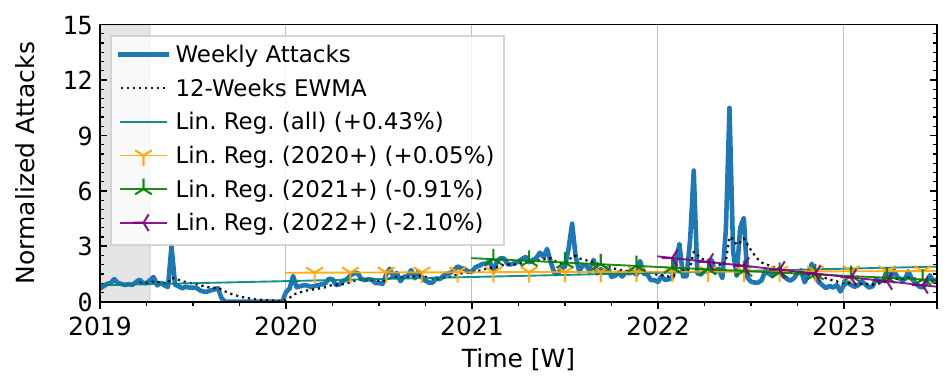}
  	\caption{Network Telescope: ORION.}
    \label{fig:normalized_attack_trends:merit}
  \end{subfigure}\hfill
  \begin{subfigure}{.5\textwidth}
    \centering
    \includegraphics[trim={0 0.2cm 0 0},clip,width=1.0\linewidth]{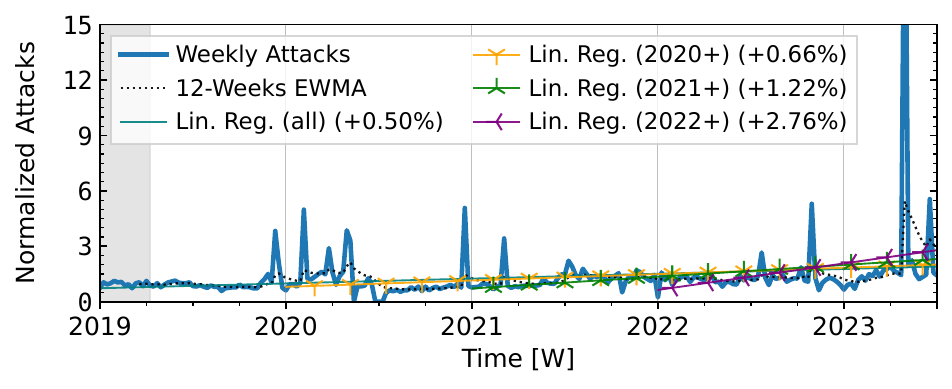}
  	\caption{Network Telescope: UCSD. The peak in 2023 reaches 27.}
    \label{fig:normalized_attack_trends:ucsd}
  \end{subfigure}\hfill
  \begin{subfigure}{.5\textwidth}
    \centering
    \includegraphics[trim={0 0.2cm 0 0},clip,width=1.0\linewidth]{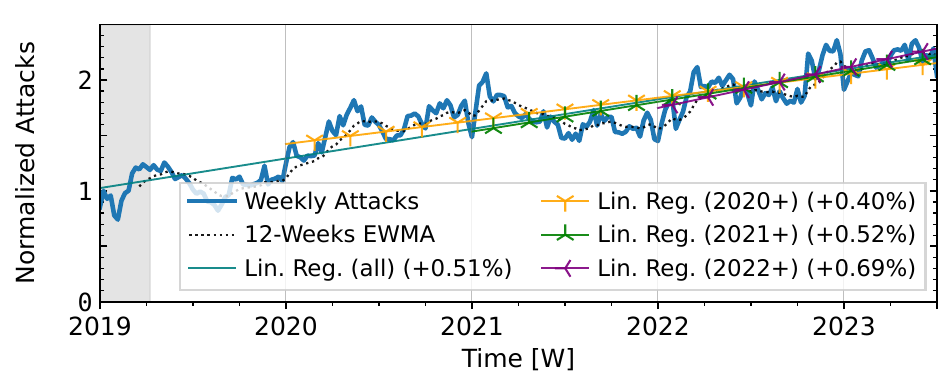}
  	\caption{Flow Data: Netscout Atlas.}
    \label{fig:normalized_attack_trends:netscoutDP}
  \end{subfigure}\hfill
  \begin{subfigure}{.5\textwidth}
    \centering
    \includegraphics[trim={0 0.2cm 0 0},clip,width=1.0\linewidth]{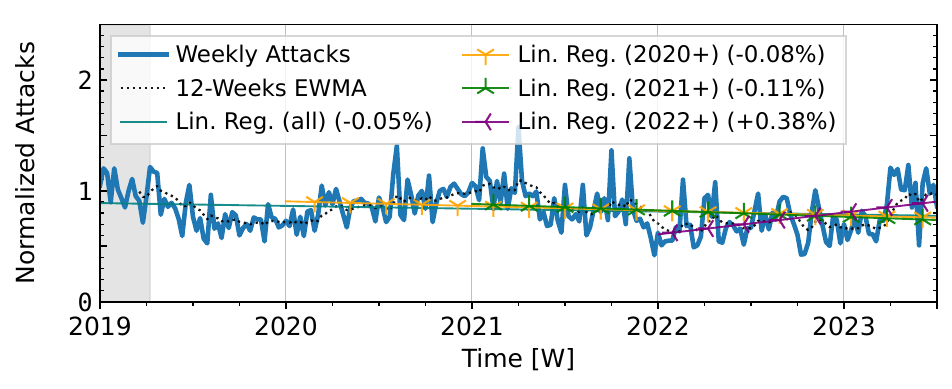}
  	\caption{Flow Data: Akamai Prolexic.}
    \label{fig:normalized_attack_trends:akamaiDP}
  \end{subfigure}\hfill
  \begin{subfigure}{.5\textwidth}
    \centering
    \includegraphics[trim={0 0.2cm 0 0},clip,width=1.0\linewidth]{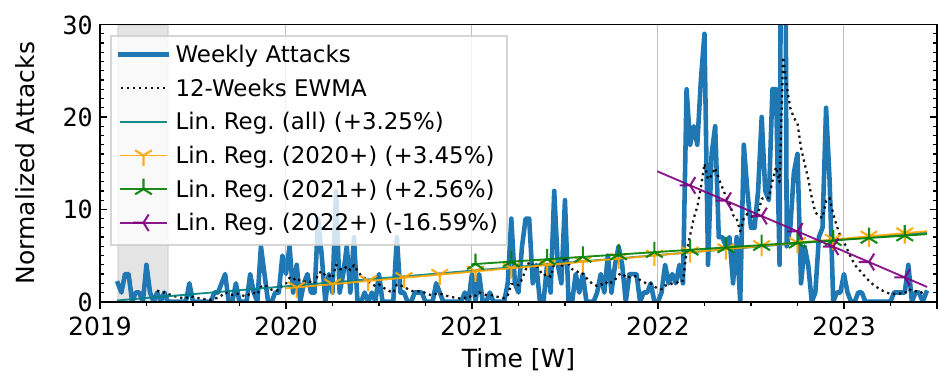}
  	\caption{Flow Data: IXP Blackholing. The peak in 2022 rises to 101.}
    \label{fig:normalized_attack_trends:ixpblackholingDP}
  \end{subfigure}\hfill
  \caption{Normalized weekly direct-path attack counts (to median of first 15 weeks as a baseline, highlighted in grey) show a growth in attacks over 4.5 years. Four observatories (ORION, UCSD, Akamai, Netscout) saw an upward trend in 2023 while one (IXP) saw a downward trend. Note y-axis scales differ.} 
  \label{fig:normalized_attack_trends:directpath}
\end{figure}

\paragraph{\inlinegraphics{emojis/honeypot.png} Honeypots observe reflection-amplification attacks} Honeypots observe DDoS attacks when their sensors are selected as amplifiers (\circled{8} in \autoref{fig:overview}).
We used data from two academic honeypots: \textit{Hopscotch}~\cite{thomas2017thousand} and \textit{AmpPot}~\cite{kramer2015amppot}. 
Although AmpPot has $\approx$70 IPs allocated, it responds from only $\approx$30, so it can associate attacks with previous scans based on which sensors it revealed~\cite{kbr-issai-16}. %

Both Hopscotch and AmpPot provided observed attack counts 
and metadata (target, length, and (estimated) packet counts).
We used algorithms developed by CCC~\cite{thomas2017thousand} to 
process both the Hopscotch and AmpPot data in the same way.
We aggregated attacks seen at multiple sensors into one event, 
including carpet-bombing~\cite{heinrich2021kids} 
against many IPs, which a single sensor may not see 
(\autoref{appendix:prefix} introduces our improved logic for 
detecting these attacks, which we shared back with CCC).

We also have access to data from \textit{NewKid}~\cite{heinrich2021kids}, but due to its single sensor (\autoref{tab:vantagepoints}) the weekly attack counts were erratic. For completeness, we include it in \autoref{appendix:newkid} but exclude it from our long-term trend analysis \autoref{sec:long-term-trends}.

\paragraph{Data aggregation}
The plots in \autoref{sec:long-term-trends} are based on attack counts for each observatory aggregated per week, \ie new attacks observed each day, summed up to weekly totals.
We normalized values to the median attack count of the first 15 weeks.
We used a similar normalization as prior work~\cite{fglpp-leicp-20} with an extended normalization period to fit the irregular nature of DDoS attacks.
This approach defines a common metric and allows data providers to keep absolute counts private.
Plots in \autoref{sec:targets} are based on distinct targets seen by each observatory per day, \ie (date, IP address) tuples.
The time series count daily tuples and sum them up to weekly totals.

\section{Comparing Long-term DDoS Trends} %
\label{sec:long-term-trends}

We now inspect the various DDoS detection data sets in detail and analyze how closely their findings correlate.
Our data aggregation is described at the end of \autoref{sec:data}.
For visualizing overall trends, we evaluated the exponentially weighted moving average (EWMA) of attack counts with a span of 12 weeks, and linear regression lines starting in 2019 to 2022 (respective slopes reported in legends).
Note that these data sets do not allow us to distinguish between missing data and the absence of attacks unless otherwise noted.

\subsection{Direct-path Attacks}
\label{sub:analysis:directpath}

\autoref{fig:normalized_attack_trends:directpath} shows normalized attack counts for observatories of direct path attacks.
(Missing data: ORION in 2019Q3-Q4, IXP in Jan 2019.)

\begin{figure}
  \begin{subfigure}{.5\textwidth} 
    \centering
    \includegraphics[trim={0 0.2cm 0 0},clip,width=1.0\linewidth]{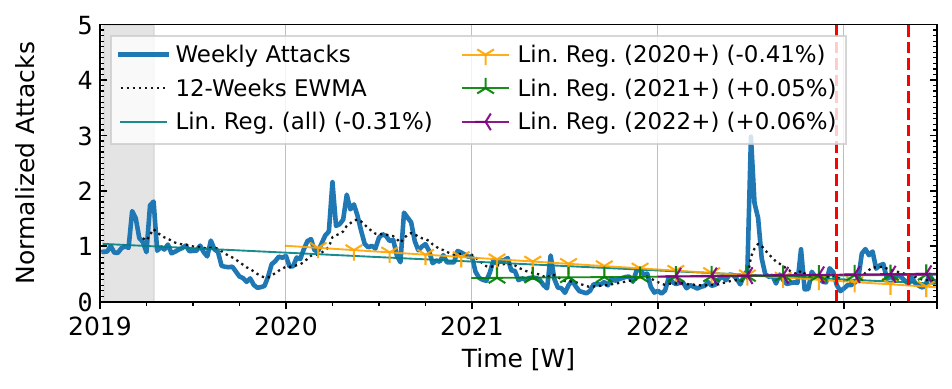}
    \caption{Honeypot: Hopscotch.} 
    \label{fig:normalized_attack_trends:ccc}
  \end{subfigure}\hfill 
  \begin{subfigure}{.5\textwidth} 
    \centering
    \includegraphics[trim={0 0.2cm 0 0},clip,width=1.0\linewidth]{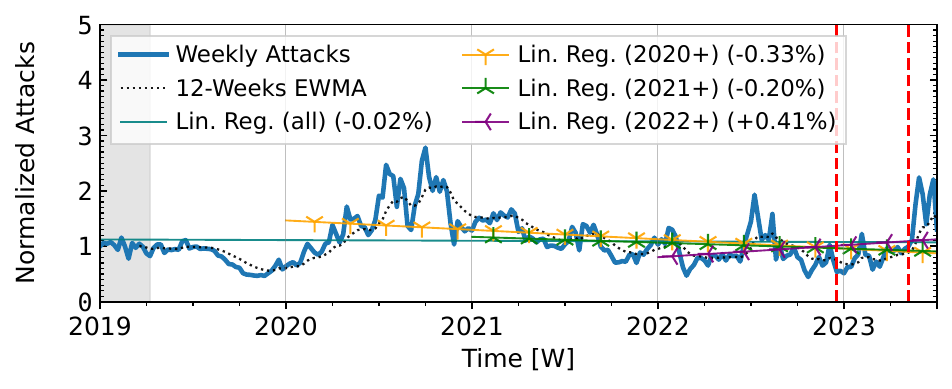}
    \caption{Honeypot: AmpPot.} 
    \label{fig:normalized_attack_trends:amppot}
  \end{subfigure}\hfill 
  \begin{subfigure}{.5\textwidth} 
    \centering
    \includegraphics[trim={0 0.2cm 0 0},clip,width=1.0\linewidth]{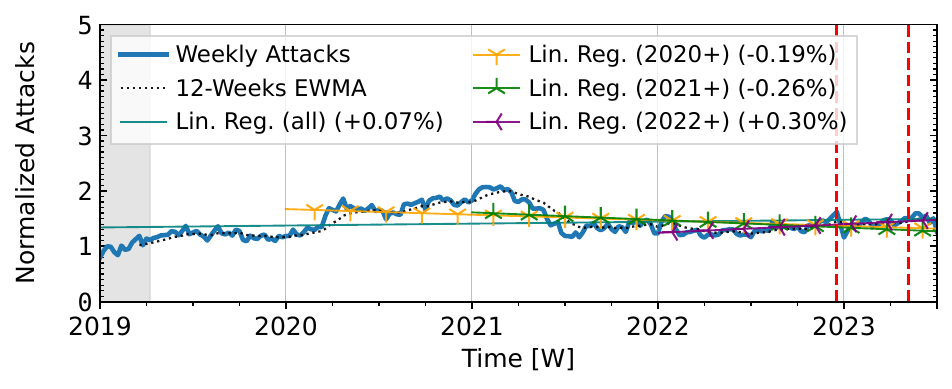}
    \caption{Flow Data: Netscout Atlas.} 
    \label{fig:normalized_attack_trends:netscoutRA}
  \end{subfigure}\hfill
  \begin{subfigure}{.5\textwidth}
    \centering
    \includegraphics[trim={0 0.2cm 0 0},clip,width=1.0\linewidth]{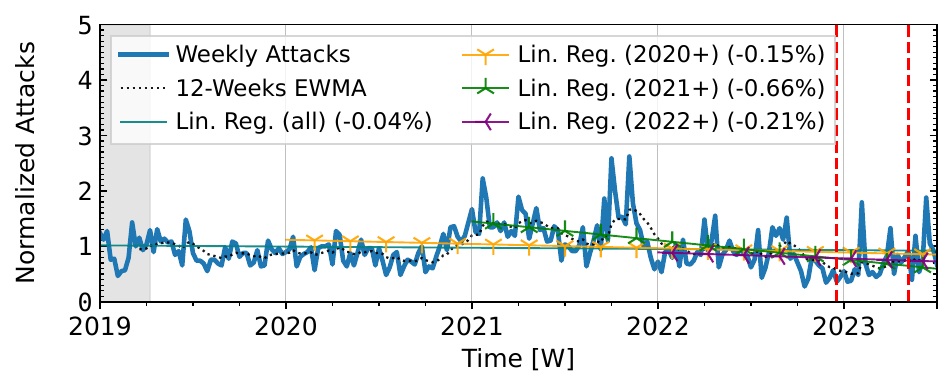}
  	\caption{Flow Data: Akamai Prolexic.}
    \label{fig:normalized_attack_trends:akamaiRA}
  \end{subfigure}\hfill
  \begin{subfigure}{.5\textwidth} 
    \centering
    \includegraphics[trim={0 0.2cm 0 0},clip,width=1.0\linewidth]{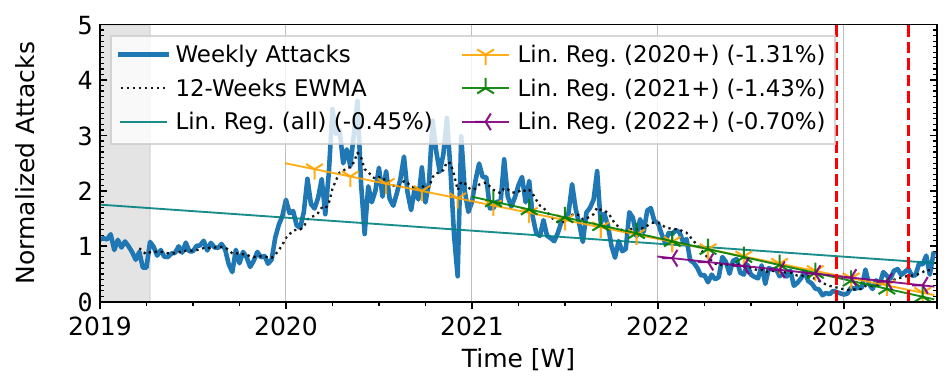}
    \caption{Flow Data: IXP Blackholing.}
    \label{fig:normalized_attack_trends:ixpblackholingRA} 
  \end{subfigure}\hfill
  \caption{Normalized weekly reflection-amplification attack counts (to median of first 15 weeks as a baseline, highlighted in grey) show varying behavior over 4.5 years. The most striking similarity is the rise in attacks in 2020, and subsequent drop across 2021. Attacks rise again in 2023,  except for Hopscotch. Red dashed lines mark DDoS takedowns by law-enforcement.} %
  \label{fig:normalized_attack_trends:reflampl} 
\end{figure}

\paragraphNoDot{The telescopes} observed a similar rise in attacks (Figures \ref{fig:normalized_attack_trends:merit}, \ref{fig:normalized_attack_trends:ucsd}).
They repeatedly saw short peaks that at least tripled attack counts; these peaks did not coincide in time.
ORION saw its largest peaks in the first half of 2022 with smaller peaks in 2019Q2 and mid-2021.
In contrast, UCSD saw its largest peak in 2023Q2 and small peaks in each year. 
In 2021 both telescopes saw an increase in attacks until summer, followed by a mild decrease until the end of the year.
Although ORION attacks peaked in 2022Q1 and Q2 the subsequent downward trend continued throughout 2022.
UCSD trends remained positive for all sub-periods.

We identified three possible reasons for these divergent views of randomly-spoofed DoS attacks.
\one The UCSD telescope is roughly 20x larger than ORION, so ORION will see fewer packets of the same attack---provided the spoofed source addresses are uniformly distributed (\S\ref{sec:data}), which makes it harder to distinguish attacks. 
\two Attacks may be too short-lived to rotate source addresses across the entire address space.
\three Some attackers might exclude known telescopes from their address selection.
Past studies showed that even telescopes of the same size will observe different IBR~\cite{wkbjh-ibrr-10} and that telescopes in topological proximity, which is not the case here, capture more similar observations~\cite{pybpp-cibr-04}.

\paragraphNoDot{Netscout Atlas} observed a relatively stable growth of attacks with an exception in 2021 (\autoref{fig:normalized_attack_trends:netscoutDP}).
Netscout presumably has a stable customer base with persistent sensors that  monitor traffic for attacks.
The first of two peaks, in 2020Q2, overlapped with peaks at UCSD and the IXP but with different dynamics and amplitudes.
The peak in 2021Q1 partially overlapped with small peaks at the IXP and Akamai.
One consistent aspect of the IXP and Netscout data (both based on observing two-way user traffic) is that relative attack counts reached a peak during the first half of the year (2019-2022) followed by a valley.
The overall trend remained upward.
However, visibility of attacks was limited to customers willing to share~data.

\paragraphNoDot{Akamai Prolexic} observations differed from other DP observatories as they
exhibited a slight downward trend over the total period (\autoref{fig:normalized_attack_trends:akamaiDP}).
Attacks detected at Akamai remained relatively stable, hence the small y-axis scale.
A valley in the second half of 2019 was followed by a high at the beginning of 2021 and a subsequent downward trend in 2021.
A rise in attacks throughout 2020 was also observed by ORION and Netscout, 
but with more pronounced amplitudes.
While peaks in 2021 rose by a factor of $\approx 1.5$ the baseline, attacks decline overall, which  was jointly observed with the IXP.
While there were several peaks in 2022, the minima dropped below $\approx$0.5$\times$ of the baseline.
Attacks rose again in 2023, but even the peaks remained below 1.3$\times$ of the baseline.

\paragraphNoDot{The IXP} observed an increase in direct-path attacks (\autoref{fig:normalized_attack_trends:ixpblackholingDP}). 
Attack counts were more erratic, often dropping to zero.
Attack counts jumped $\approx$10$\times$ from their baseline in the first half of 2020 and 2021, and $\approx$30$\times$ in 2023, but dropped for the second half of each year.
The closest overlap with the telescopes was the higher activity in mid-2021, which was only a small elevation ($\approx$2$\times$) in the UCSD data.
An increase in activity was common to UCSD in 2020Q1/Q2 and ORION in 2022Q1/Q2.
Note the IXP data reflects traffic for which customers requested blackholing.
It is thus a lower-bound of direct-path attacks passing this~IXP and may depend on IXP customer actions.

\paragraph{Trends in direct-path attacks} 
The linear regressions show that four of the five observatories experienced an upward
trend over the full measurement period.
A few peaks correlate across multiple data sets, albeit at different amplitudes.
In 2023, three observatories saw a slight upward trend (UCSD, Akamai, Netscout) while ORION saw 
no trend, and the IXP saw a downward trend.
These divergent observations across vantage points reflect the limited and disparate
coverage of each sensor instrumentation (\S\ref{sec:targets}).

\subsection{Reflection-amplification Attacks}
\label{sub:analysis:reflampl}

\autoref{fig:normalized_attack_trends:reflampl} shows the evolution of reflection-amplification attacks in our data sets.
As in \autoref{fig:normalized_attack_trends:directpath}, the y-axis range shows the difference in observed attacks as a factor of the normalized week count.
We marked dates of known DDoS takedown operations with red dotted lines in these plots.
Per seizure warrants, these happened on Dec~13, 2022, and May~4, 2023~\cite{usdc-swtor-23}.

\paragraphNoDot{The honeypots} in our study observed a pronounced growth of attacks in 2020 (\autoref{fig:normalized_attack_trends:ccc},\ref{fig:normalized_attack_trends:amppot}).
Hopscotch recorded most attacks early in 2020, when Netscout and IXP counts also increased but with different relative amplitudes.
In contrast, AmpPot saw its highest attack peaks later in 2020, mysteriously when Hopscotch peaks declined.
Notably, all honeypots (HP) observed a decline in attacks from late 2021 until mid-2022, when both observed a spike not visible at the industry observatories (Netscout, Akamai, IXP). 
This downward trend is consistent with industry data (\autoref{fig:normalized_attack_trends:netscoutRA}, \ref{fig:normalized_attack_trends:akamaiRA},\ref{fig:normalized_attack_trends:ixpblackholingRA}) and industry efforts to deploy SAV (see discussion in \autoref{subsec:history}).

\paragraphNoDot{Netscout Atlas} exhibited a stable, mild upward trend with a pronounced rise until 2021, quick declines in early 2021 and early 2022, and a slow rise starting in Q1 of 2022 (\autoref{fig:normalized_attack_trends:netscoutRA}).

\paragraphNoDot{Akamai Prolexic} saw only small variations in attacks until 2020Q3, when attacks increased with a peak above $2\times$ its baseline in 2021Q1 (\autoref{fig:normalized_attack_trends:akamaiRA}).
The subsequent peak in 2021Q1 coincides with a period of high attack
counts for Netscout, the IXP, and AmpPot.  Like others, Akamai saw a
decrease in attacks in the first half of 2021.  However, the peaks in
2021Q4 are unique to Akamai.  After dropping to $\approx$0.5$\times$ in
late 2022, attacks increased again.

\paragraphNoDot{The IXP} 
observed a slow decline of attacks through most of 2019, followed by a steep increase until 2020Q2 (\autoref{fig:normalized_attack_trends:ixpblackholingRA}).
Hopscotch and Netscout similarly observed a rise in attacks during 2020Q2,
although  at lower amplitudes.  2020Q4 was a second period of high attacks
with a subsequent decline that continued (punctuated by bursts of attacks)
until 2023.  While the decrease in 2021Q2 was also observed by Netscout
and Akamai, attack counts at the IXP decreased until the turn to 2023.  After
a low at the turn to 2023, attacks increase again but stayed below
the baseline.  This time series was more stable than IXP direct-path
attacks (\autoref{fig:normalized_attack_trends:ixpblackholingDP}).

\begin{figure*} 
   \center
    \includegraphics[width=1.95\columnwidth]{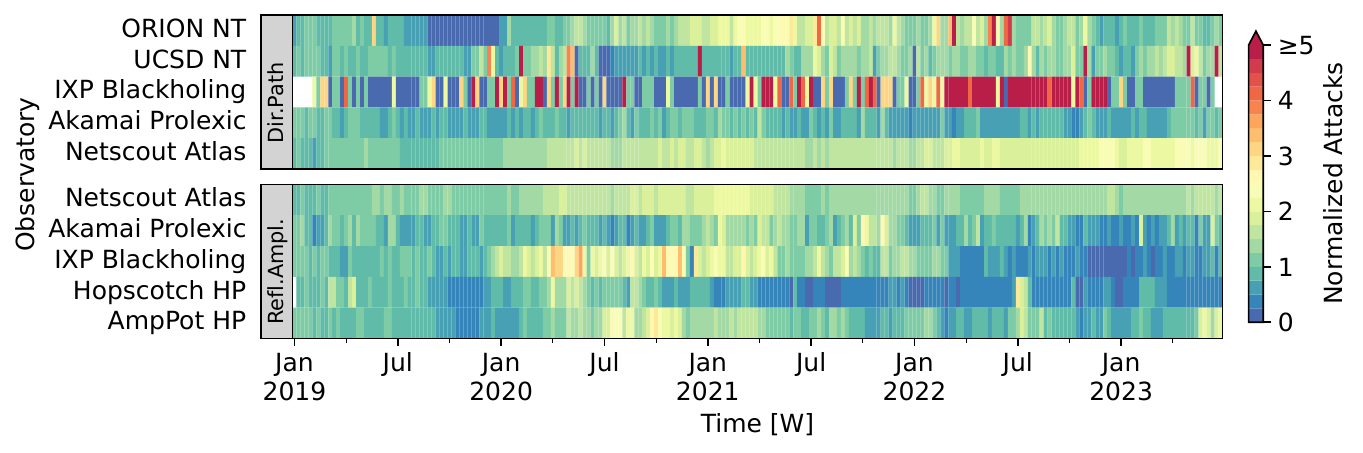}
	\caption{Normalized weekly attack counts observed at our 10 vantage points. Direct-path (DP) attacks (top 5 rows) increased in 2022 while reflection-amplification (RA) attacks (bottom 5 rows) had highest intensities during 2020 and declined~thereafter.} 
    \label{fig:mutual_heatmap_atk_norm} 
\end{figure*}

\paragraph{Trends in reflection-amplification attacks}
No pair of time series exhibits similar behavior for the whole period, but all five vantage points showed the increase in attacks in 2020 followed by a decrease 2021.
While Akamai, the IXP, and AmpPot saw this downward trend continue through 2022, Netscout and Hopscotch saw a flatter trend that year. 
Finally,  attacks rose again through 2023 except for Hopscotch, which only saw a short peak in Q1.
There were also short periods (3-6 months), in which two or more time series proceeded similarly, \ie 
\one Hopscotch, AmpPot, IXP 2019Q4,
\two Hopscotch, Netscout, IXP in 2020Q2,
\three AmpPot, Netscout, Akamai, IXP (dip in mid 2021), 
\four Hopscotch and AmpPot (peak mid-2022),
and \five all observatories had a low in January 2023.

\begin{figure}
  \center
  \includegraphics[width=1.0\columnwidth]{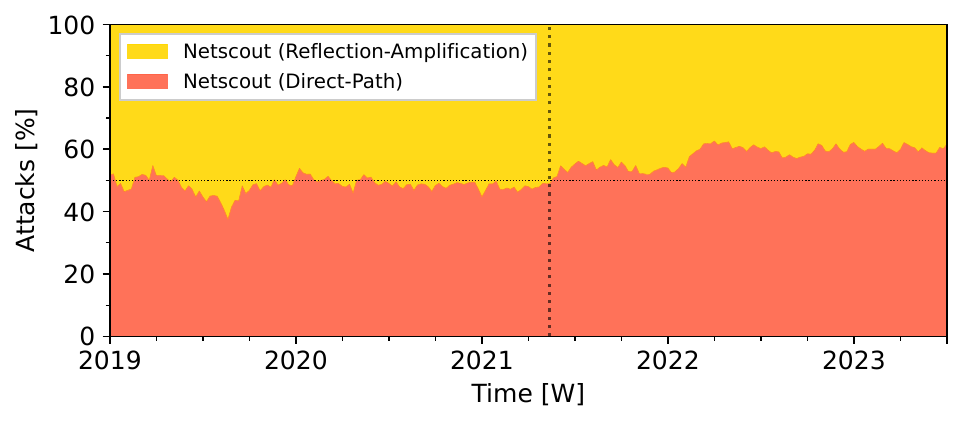}
  \caption{The relative share of reflection-amplification (RA) and direct-path (DP) attacks observed by Netscout per week shows a shift toward DP attacks. The horizontal line indicates 50\%, \ie equal share. The dotted vertical line marks the latest crossing of the 50\% mark.}
  \label{fig:shift:netscout}
\end{figure}

\paragraph{DDoS-service takedowns by law enforcement}
Arrests and infrastructure seizures should have an immediate effect on attacks~\cite{BooterShutdown}.
Two DDoS-takedown efforts during our observation time left an indeterminate footprint.
The first in late 2022Q4 was followed by immediate, (small) valleys at the turn to 2023 in all four graphs.
In contrast, the takedown in 2023Q2 was only followed by valleys in \autoref{fig:normalized_attack_trends:akamaiRA}, \autoref{fig:normalized_attack_trends:ixpblackholingRA}, and \autoref{fig:normalized_attack_trends:ccc}.
Even if these changes were caused by takedowns, their impact on DDoS trends remained insignificant in our time series.

\subsection{Trends Across Attack Types}
\label{subsec:trends_across_attack_types}

We now take a comparative view of trends across all attack types, even
though the different methods and coverage in measurements make this
challenging.

\paragraph{Shifts between attack types} \label{ssub:relativeshift}
\autoref{fig:mutual_heatmap_atk_norm} combines the time series 
from \autoref{fig:normalized_attack_trends:directpath} and \autoref{fig:normalized_attack_trends:reflampl} into a single heatmap. 
Colors represent normalized attack counts.
We arrange observatories by attack type: direct-path (top) and reflection-amplification (bottom).
Overall, we find that time series data of the same attack type---with the exception of Akamai---are more positively correlated, but periods of anti-correlation also exist.
Most direct-path attacks trended toward increased attack counts in early 2022, 
although at different intensities, before observation diverged.

In contrast, Akamai saw higher attack counts during 2019 and 2020 followed
by a downward trend until 2023.  Reflection-amplification attacks showed
higher intensities between 2020Q2 and the end of 2021Q2 but lacked a
clear trend toward the end.  Again, Akamai differed as attack counts
increased later with the highest peak in 2021Q4, and continuously show
smaller peaks throughout 2022.  Even for the same attack type and
measurement method, we can see opposite trends, such as for AmpPot and
Hopscotch in 2023.

When comparing across observatory types, coherence lowers as the observed attack types differ. %
Netscout, which measures both direct path and reflection-amplification attacks at the same platform, observes a relative shift toward direct path attacks based on absolute attack counts (\autoref{fig:shift:netscout}).
The dotted line marks the shift in 2021Q2, which matched the downward trend of RA attacks (\autoref{fig:normalized_attack_trends:netscoutRA}).
This roughly echoes our observation (\autoref{fig:mutual_heatmap_atk_norm}) that RA attacks were relatively high in the first two years while direct-path attacks increased relatively in the latter years.
In contrast, Akamai consistently reported a larger share of direct-path attacks throughout the entire period.

\begin{figure*} 
   \centering
    \includegraphics[width=0.9\linewidth]{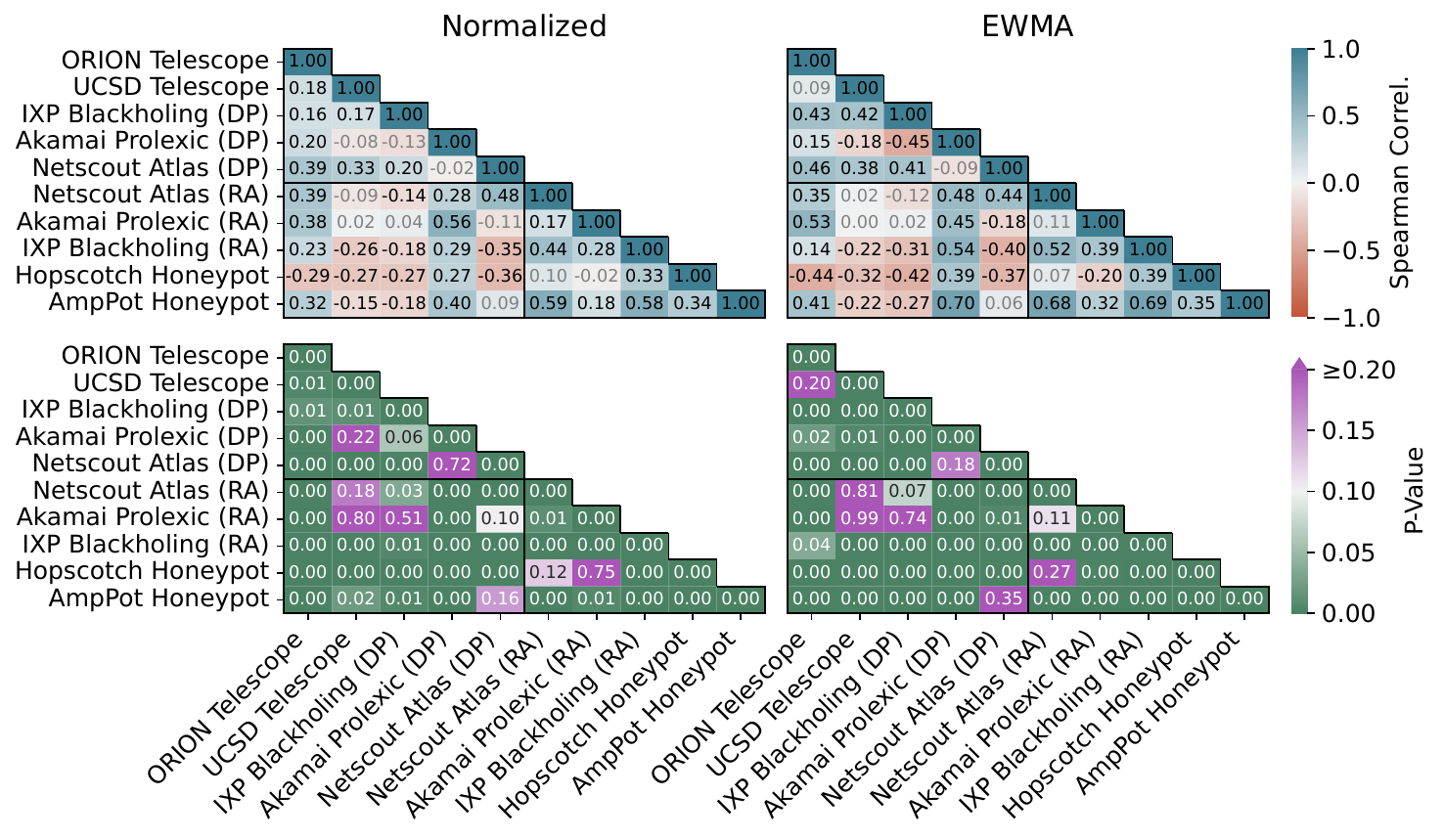}
	\caption{\textit{Spearman correlation}: Platforms that observe the same kind of DDoS event (direct-path or reflection-amplification) show higher correlation. The Akamai direct-path time series is an exception as it correlated with reflection-amplification observatories. The top two graphs show the Spearman correlation for the normalized data on the left and the EWM on the right. A grey font marks correlation coefficients with a p-values above 0.05. The bottom graphs show the respective p-values.} 
   \label{fig:vp_correlation:spearman} 
\end{figure*}

\paragraph{Correlations between attack trends}
\autoref{fig:vp_correlation:spearman} shows Spearman correlations between pairs of observatories--a linear value between 1 (correlation) and -1 (anti-correlation).
We applied the Spearman correlation because it calculates a monotonic correlation that is less susceptible to outliers than Pearson, a linear correlation.
We calculated correlations for the normalized data (left) and the weighted moving average (EWMA) (right).
Correlations with p-values (bottom part of the figure) above 0.05 are considered statistically insignificant and have their font greyed out.

We found a low to modest correlation between Netscout (DP) and direct-path (DP) attack observatories (ORION 0.2, UCSD 0.33, IXP (DP) 0.39) -- except Akamai (p-value $>$0.05).
Akamai (DP) showed a low correlation with ORION (0.20), but exhibited high p-values in correlations with other direct-path time series (0.06-0.72).
Other DP observatories correlated weakly (0.16-0.20).
Correlations between the EWMA were more pronounced.
ORION and IXP (DP) showed the same medium correlation as Netscout and other observatories (0.43).
The correlation between UCSD and ORION was statistically insignificant (p-value: 0.2).
Akamai (DP) was anti-correlated with UCSD (-0.18) and the IXP (DP) (-0.45).

Our RA observatories showed higher correlations than the DP observatories, with low to moderate pairwise correlations (0.17 to 0.59).
Hopscotch had statistically insignificant correlations with Netscout (RA) (p-value: 0.12) and Akamai (RA) (p-value: 0.75).
The EWMA lines correlated more strongly (0.26-0.69) when significant.
For EWMA, Hopscotch and Akamai (RA) had a low p-value and exhibit an anti-correlation (-0.20).
The p-value of Akamai (RA) and Netscout (RA) increased leaving the correlation insignificant.

Two observatories stood out in correlations across attack types: \one Akamai (DP) showed positive correlations with observatories of the opposite (RA) type (0.27-0.56).
\two ORION showed positive correlations with four of the five RA observatories: Netscout (RA) (0.39), Akamai (RA) (0.38), IXP (0.23), and AmpPot (0.32).
Additionally, the two Netscout time series had a medium correlation (0.48).
Analysis between observatories showed low to medium anti correlations (-0.14 to -0.36).
In addition to two positive correlations with ORION and Akamai (DP), the Akamai (RA) time series only had statistically insignificant correlations with the other three direct-path observatories.
Two more correlations were statistically insignificant: Netscout (RA) \& UCSD and AmpPot \& Netscout~(DP).

Apart from Akamai, time series of the same attack type tended to correlate more strongly, \ie the attack counts evolved in similar ways.
This correlation did not hold for all pairs of observatories within each group, implying that different observatories of the same group do not uniformly observe the same attack events---consistent with our earlier findings.
While our trend lines were more consistent within the direct-path group the correlation analysis shows higher values for the RA observatories.

We cross-checked our results by calculating the Pearson correlation.
It confirmed our results, even if correlation values varied slightly: correlation was stronger among RA observatories and weaker among DP observatories.
\autoref{appendix:correlation:quarterly} provides quarterly pairwise correlations and similarly shows stronger correlations among observatories that observe the same attack type.

\paragraphNoDot{Why does Akamai see different trends?}
The data from Akamai Prolexic represents attack events from traffic that transited its AS.
Customers must own a prefix that can be rerouted through the Prolexic AS for attack mitigation.
This requirement will affect attack methodologies and trends in their data.
Netscout and the IXP also observe with biasing characteristics.
This reality is an inherent challenge to understanding a competitive private sector 
landscape with no standardized reporting of the analyzed phenomena.

\paragraph{Pandemic}
As lockdowns drove people to spend more time online~\cite{fglpp-leicp-20}, DDoS attacks increased and diversified~\cite{bkn-ddsap-23,fola-emdad-24}.
The observatories in our study saw an increase in both DP (\autoref{fig:normalized_attack_trends:ucsd},\ref{fig:normalized_attack_trends:ixpblackholingDP}, \ref{fig:normalized_attack_trends:akamaiDP}, \ref{fig:normalized_attack_trends:netscoutDP}) and RA (\autoref{fig:normalized_attack_trends:netscoutRA},\ref{fig:normalized_attack_trends:ixpblackholingRA}, \ref{fig:normalized_attack_trends:ccc}, \ref{fig:normalized_attack_trends:amppot}) attacks during 2020.

\section{Analyzing DDoS Targets}
\label{sec:targets}

Our trend analysis (\autoref{sec:long-term-trends}) 
showed rough similarities within attack types.
We now examine how targets of DDoS appeared across observatories.
During a 4.5-year observation period,  IP addresses can change owners while others rotate due to dynamic assignment.
We used the tuple \textit{(attack start date, target IP address)} (see~\autoref{sec:data}) to identify a target unless otherwise noted, and deduplicated the resulting set, obtaining   
 \num{28474161} distinct targets or \num{14565588} distinct IP addresses.

\subsection{Attackers Often Chose One Attack Type}
\label{subsec:Attackers_Usually_Chose_One_Attack_Type}

\begin{figure} 
   \centering
    \includegraphics[width=1.0\linewidth]{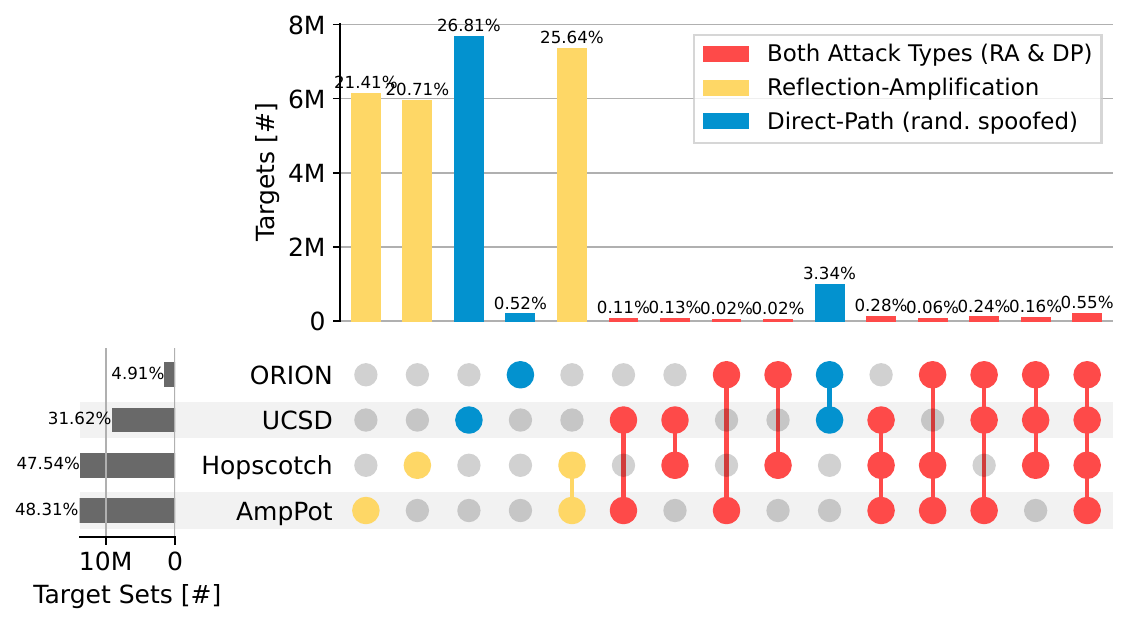}
	\caption{The UpSet plot displays targets that were exclusively seen in intersection sets of observatories on the same day. Colors signify the attack types in each set. Observatories of the same type (RA, DP) see large shares of overlapping targets. Only 0.55\% of targets were observed by all four observatories.}
   \label{fig:targets:overview} 
\end{figure}

We first compared targets across observatories to answer the two questions:
\one How many distinct targets did each observatory see?, and
\two how large was the target overlap between all four observatories?
The UpSet plot~\cite{lhsvp-uvis-14} in \autoref{fig:targets:overview} answers both questions.
It shows the number of distinct targets (date, IP address) seen by each observatory in the left bar plot.
These shares are not exclusive, \ie they sum up to more than 100\% of distinct targets.
The top bar plot quantifies the target intersections, \ie the targets exclusively observed by the combination of observatories marked in the matrix below.
Each bar shows absolute counts and additionally displays its share among all distinct targets at the top.

Both honeypots (HP) saw nearly the same number of targets (left bars: roughly 48\%). 
ORION saw an order of magnitude fewer targets than the HPs and six times fewer than UCSD, which monitors 22 times more IP addresses.

Hopscotch and AmpPot each uniquely observed $\approx$21\% of all targets and another 25\% together (three yellow bars in the top bar plot), more than the respective individual shares.
UCSD had the highest share of uniquely observed distinct targets ($>$26\%).
The limiting factor for a larger overlap is the low target count observed by ORION.
Except for UCSD (14\%) the share of overlapping targets across observatories of the same kind (NT: UCSD \& ORION, HP: AmpPot \& Hopscotch) was over 50\% (AmpPot 57\%, Hopscotch 56\%, ORION 87\%).
AmpPot, for example, shared 57\% of the targets it observed with Hopscotch.
For reflection-amplification attacks, this means that attackers likely selected reflectors from multiple HPs.
Still, both honeypots observed a significant share of exclusive targets.
Randomly-spoofed DoS attacks that produce enough packets to be visible in ORION are likely to produce enough packets to appear in larger telescopes.
But even with its size, UCSD NT did not observe all targets.

Nawrocki et al.~\cite{nkkhs-advmd-23} examined the effect of different attack definitions used by honeypots and found %
a 15\%--45\% difference in attack targets.
Our data sets are based on already processed traffic data, which does not 
allow us to make statements about the effect of attack detection thresholds.

\begin{figure}
   \centering
    \includegraphics[width=1.0\linewidth]{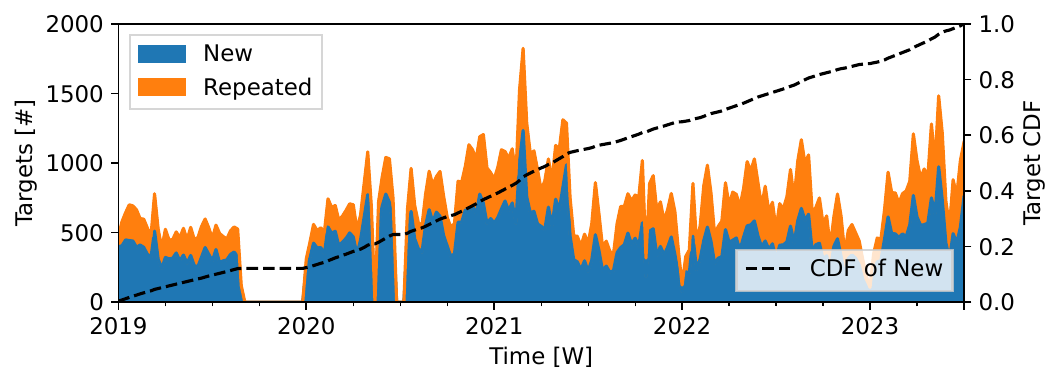}
	\caption{Target~tuples (date, IP address) observed by all four observatories (ORION, UCSD, Hopscotch, AmpPot) during our measurement period, summed up per week. These targets reflect the 0.55\% right-most set in \autoref{fig:targets:overview}. The dashed line counts new targets as CDF on the right~axis.}
   \label{fig:targets:all} 
\end{figure}

\paragraph{Highly-visible targets}
A small fraction of targets (1.57\%) was affected by multiple attack types, a third of these (0.55\% or \num{155663} targets or \num{97470} distinct IP addresses) at all four observatories. 
\autoref{fig:targets:all} shows the time series of these 0.55\% targets as a stacked area plot (left axis).
The blue area signifies new targets, \ie IPs seen for the first time, while the orange area counts recurring targets.
The dotted line plots target growth over time as a CDF (right axis).
The four observatories continuously saw new shared DDoS targets, most of which appearing between 2020Q4 and 2021Q2.
This time series does not resemble any individual trend line (\autoref{fig:normalized_attack_trends:directpath} and \ref{fig:normalized_attack_trends:reflampl}), but 
honeypots saw periods of higher attack counts during these times.
Leaving out the relatively small ORION, the three remaining observatories saw an additional 0.28\% overlapping targets (50\% more).
This still adds up to fewer than 1\% of all targets (\autoref{fig:targets:overview}).

The largest share of highly-visible targets belonged to OVH (AS16276, 18.8\%), followed by Hetzner (AS24940, 5.1\%) and Amazon (AS16509, 2.69\%).
Including these three, 7~of our top 10 most targeted ASes belong to hosters.
Hosters usually offer DDoS-protection-as-a-service, which may lead attackers to use multiple attack vectors, \eg RA and DP attacks, to overcome defenses.

In a two-year study (2015-17) Jonker et al.~\cite{jkkrs-mtuam-17} examined the overlap between AmpPot and UCSD NT.
They found 4.5\% (282k) shared IP addresses, half of which were hit by attacks at the same time.
In our data, this overlap is lower, \ie 1.18\%--2.9\%  of the IP addresses.
Jonker et al.~also saw OVH as the main target (12.3\%), followed by China Telecom (14th largest share of targeted IPs in our data) and China Unicom/AS4837 (7th).
See \autoref{appendix:overlap-asn} for the top 10 ASes by number of highly-visible targets.

\subsection{Target Overlap with Industry}
\label{sec:overlap-indsutry}

\begin{figure}
  \centering
  \includegraphics[trim={2.8cm 0 0 0},clip,width=1.0\linewidth]{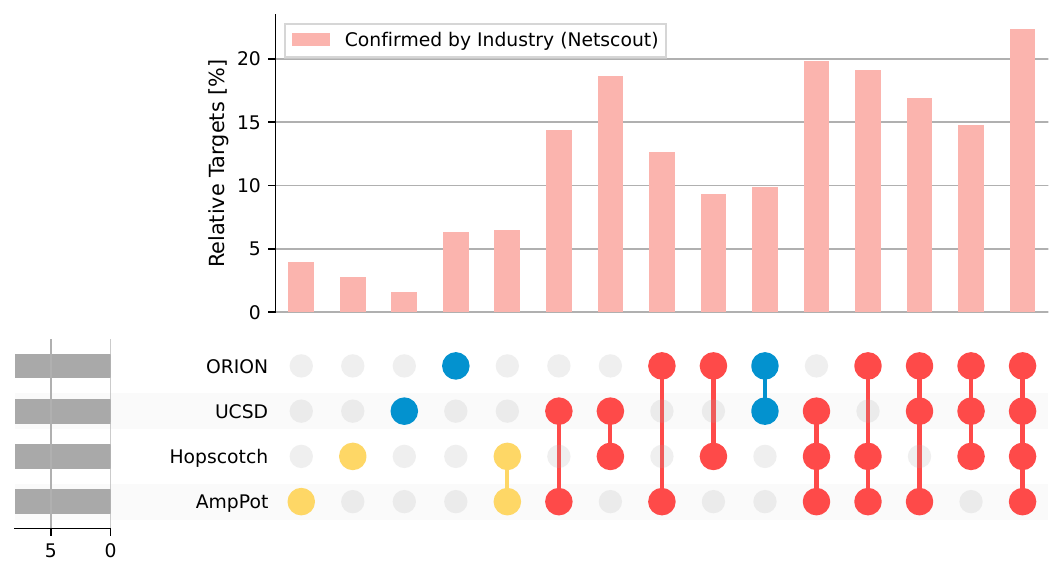}
  \caption{Relative share of targets observed by academia and confirmed by industrial baseline data from \textit{Netscout}. Supersets are shown in \autoref{fig:targets:overview}. Netscout baseline data shows the largest relative overlap with the targets seen by all four observatories. These are likely large, multi-vector~attacks.}
  \label{fig:industry:netscout}
\end{figure}

\paragraph{Netscout}
We extended the target overlap analysis to industry with a focus on federated attack inference.
Netscout compared targets from academia (\autoref{fig:targets:overview}) to their baseline data set---constituting approximately 28\% of all Netscout alerts.
The shares of confirmed targets are presented in \autoref{fig:industry:netscout}.
There are important caveats when comparing Netscout and academic attack observations.
First, Netscout's anonymized data limits complete confirmation.
Second, Netscout excludes attack alerts below the product-defined ``medium'' threshold, which may prevent confirmation of less severe attacks seen by academic observatories.
Lastly, Netscout observations are made at intermediate systems on network paths of customers that contribute alert feedback as opposed to the endpoint nodes that are party to all attack events observed.
Netscout targets had a 2\%-6\% overlap with each \emph{individual} observatory, but a 20\% overlap with the set of targets seen by all academic observatories, \ie 20\% of the 0.55\% in \autoref{fig:targets:all}.
Larger, multi-vector attacks were more likely seen from all vantage points.

\begin{figure*}
  \begin{subfigure}{.49\textwidth} 
    \centering
    \includegraphics[width=0.98\linewidth]{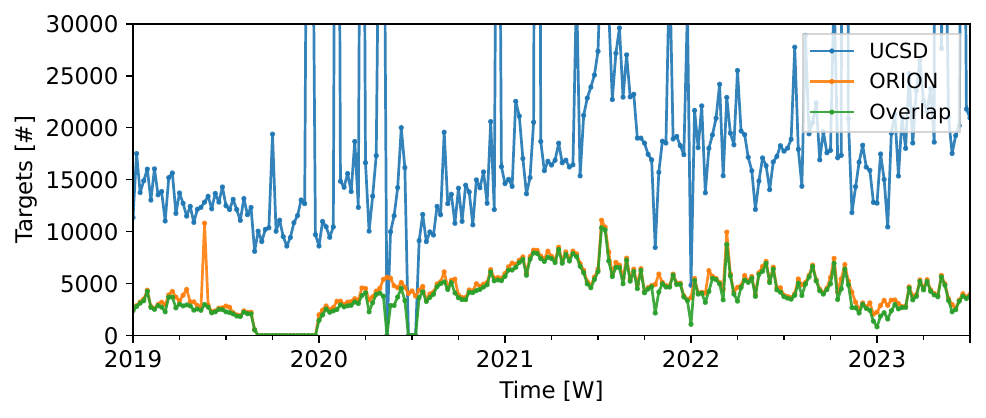}
    \caption{The UCSD telescope observed most targets seen by ORION.}
    \label{fig:targets:telescopes} 
  \end{subfigure}\hfill
  \begin{subfigure}{.49\textwidth} 
    \centering
    \includegraphics[width=0.98\linewidth]{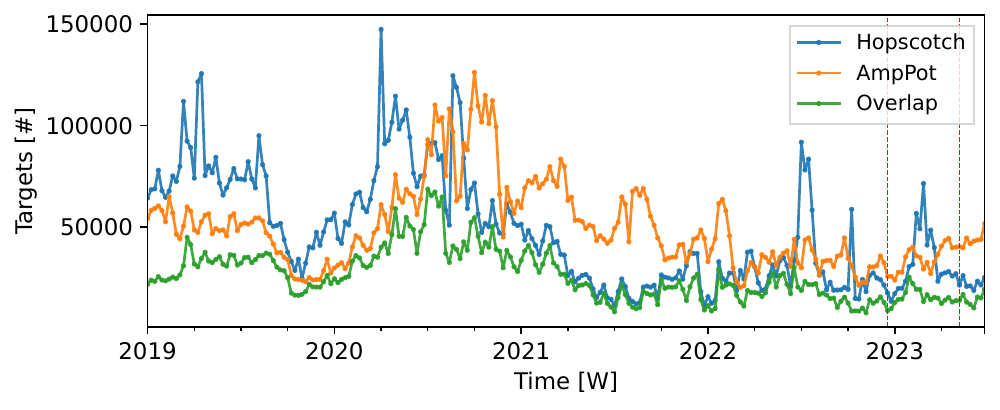}
    \caption{Both honeypots uniquely observed a large share of targets. }
    \label{fig:targets:honeypots} 
  \end{subfigure}\hfill
	\caption{Weekly observed targets (unique per day) for telescopes (a) and honeypots (b) and respective overlapping observations.}
  \label{fig:targets:timeseries} 
\end{figure*}

We assessed how many targets inferred by Netscout were also observed by academia.
With 23\% of the baseline data set this is a substantial but partial view.
No academic observatory independently saw all shared targets, 
with overlaps of 15.2\%, 13.6\%, 5.7\%, and 3.1\%, respectively.
This intersection analysis compares attack type-specific observatories with \emph{all} DDoS attacks as observed by Netscout, which generally will result in lower intersections.

\paragraph{Akamai}
The overlap analysis with targets in the network prefix of Akamai showed minor overlaps ($<$0.25\%), about $100\times$ lower than Netscout.
See \autoref{appendix:targets:akamai} for a detailed plot.
A small overlap is consistent with the focus of the Akamai dataset on their ASN, which advertises a subset of prefixes on the Internet.
Together, academic research observatories saw 33\% of the Akamai target set.
Both honeypots saw a larger share of the targets ($\approx$20\%) than the respective telescopes ($\approx$7\%).
These findings underscore the value of federated DDoS inference.

\paragraph{Previous work}
Nawrocki et al.~\cite{nkkhs-advmd-23} found a 3\% overlap in targets between Hopscotch (RA) and a commercial DoS mitigation provider (RA) (Nov 21 and Jan 22).
\autoref{fig:industry:netscout} shows a similar $\approx$3\% overlap between Hopscotch and Netscout, which is a lower bound as mixed attack type shares are not considered.
A 7-month study in 2019-20~\cite{kdh-dndip-21} found 33\% overlap of targets %
between IXP blackholing and honeypot observations.

\subsection{Target Overlap over Time}

\autoref{fig:targets:timeseries} shows the overlap 
in observed targets per day (summed up per week)
between telescopes and honeypots. 
Blue and orange are the respective observatories 
while the green line marks the overlapping subset.

\paragraph{Telescopes}
The telescopes observed 32.23\% of all targets, 95\% (30.66\%) of these 
{\em only} the telescopes observed \autoref{fig:targets:telescopes}.
Matching the observation from \autoref{fig:targets:overview}, UCSD observed most targets seen by ORION.
Unlike their trend graphs (\autoref{fig:normalized_attack_trends:merit}, \ref{fig:normalized_attack_trends:ucsd}), their target lines show similar behavior: a downward trend towards 2020, an upward trend until 2021Q3, and a low point at the turn of 2023 followed by a small upward trend.
The UCSD line has many more spikes.
The high in mid-2021 also appears in the attack trend graphs (\autoref{fig:normalized_attack_trends:merit}, \ref{fig:normalized_attack_trends:ucsd}). 

\paragraph{Honeypots}
Both honeypots (HP) saw the same order of magnitude of targets (\autoref{fig:targets:honeypots}).
Their observations add up to 69.33\% of all unique targets and 67.76\% of targets they observed exclusively.
While the shared targets were usually a subset of both HP target sets, there was a period from mid-2021 to mid-2022 where AmpPot observed most of the targets seen by Hopscotch.

The target graphs show a similar behavior to the respective attack trends (compare \autoref{fig:normalized_attack_trends:ccc}, \ref{fig:normalized_attack_trends:amppot}).
We marked the DDoS takedowns once more with red dotted lines. %
All observatories saw a small valley around the first takedown, followed by a rise in targets.

An attacker must select a honeypot as a reflector for that 
honeypot to observe the attack.  Differences in protocol support across honeypots
will affect the composition of attacks they see.
As an example, AmpPot observed more targets attacked via CHARGEN while Hopscotch saw more targets attacked via CLDAP until mid-2020.
For protocols such as QOTD, RPC, and NTP both had largely overlapping target sets.

\vspace{-0.3cm}
\section{Related Work}
\label{sec:related-work}

The last decade has yielded a vast range of DDoS studies, of which we can only present a snapshot here.

\paragraph{Attack characterization}
Studies have characterized DDoS attacks~\cite{simra2023ddos2vec,jkkrs-mtuam-17,kdh-dndip-21}
as well as abusable protocols~\cite{khrh-hhatr-14,nawrocki2021quic,nthms-ibtcq-22,r-ahrnp-14,
kk-nr-14,tkjrr-amadp-21,otsw-fbtfy-22,akpa-opai-17,nawrocki2021ecosystem}.
They quantified the state of amplifier deployment~\cite{nawrocki2021transparent},
identified scan infrastructure~\cite{kbr-issai-16},
and booter services~\cite{dbpa-ladab-17},
measured the adoption of DDoS services~\cite{jonker2016protection}
enabling flow telemetry-based traceback~\cite{kr-babta-21},
examined detection methods~\cite{bbcs-dzaup-20,nmaa-epddd-22,mas-dcdsg-22},
quantified attacks~\cite{thomas2017thousand,bgd-qsdai-17,gd-qtsdr-20,jkkrs-mtuam-17,gd-wtmc-2018}, 
 discovered new attack vectors~\cite{b-ccdda-19,kopp2019hide,mhsbw-idmls-20,kdh-dndip-21,nawrocki2021ecosystem,mchh-temvd-21,bafhw-wmtra-21}, and
criminal techniques, tactics, and procedures~(TTPs)~\cite{kopp2019hide,griffioen2021adversarial,hnkds-sunws-22,hnsw-rvmls-22}.
Few research efforts moved beyond a single class of attacks~\cite{jkkrs-mtuam-17,jonker2018-blackholing} or arrived at insights by studying the overlap between independent datasets \cite{nkkhs-advmd-23}.
To our knowledge, we provide the first comprehensive, longitudinal trend analysis of the two most concerning DDoS attack types, enabled by  vantage points from research and~industry.

\paragraph{Mitigation}
Prior studies have focused on proactive measures and reactive responses. 
These include the efficacy of operational controls, such as blackholing \cite{hnjds-pfdms-18,ixpscrubber,nawrocki2019down,jonker2018-blackholing}, traffic engineering \cite{rbch-aanpf-22}, or anycast \cite{mshvm-adenr-16}.
Other work measured the adoption of protection techniques and resulting resilience \cite{jonker2016protection,rbch-aanpf-22}, and explored mitigation options in improved protocol design \cite{r-ahrnp-14}, collaborative detection and response \cite{kr-babta-21,DXP}, and law enforcement intervention \cite{kopp2019hide,BooterShutdown,moneva_effect_2023}. 
Our work analyzed attack trends, which informs where to focus mitigation techniques. 

\pagebreak

\vspace*{-0.36cm}
\paragraph{Open challenge}
A graphical taxonomy of recent related work is presented in
\autoref{appendix:relwork}. While each study contributes to the overall
DDoS landscape, our community has not yet arrived at a common and
comprehensive understanding of DDoS.  We aim to narrow this gap with our
data-driven analysis involving multiple communities, and advocate steps
to support a science of DDoS assessment in the future.

\vspace{-0.2cm}
\section{Implications for Public Policy} 
\label{sec:future-ddos}

Persistence and prevalence of DDoS poses a systemic risk to the
ecosystem, and researchers and operators do not understand the current
extent of harms or effectiveness of mitigations.  The breadth
and scope of this study suggests the inherent limits to what
the academic community can provide on its own.

Growing awareness of the obstacle of opacity has motivated 
governments to advocate for more transparency.  The EU
NIS2 directive (2021) requires essential and important entities
to notify their competent authority of any incident that
``significantly impacts'' provision of their 
service \cite{nis2-eu-2021}.
Similar U.S. regulations are in discussion \cite{cisa24}. 
These developments offer an opportunity
to inform consideration of how mandated data sharing
of DDoS incidents could occur to ensure its utility 
to the regulatory objective.  

In particular, mandated reporting does not usually
involve publishing the information or even sharing it with independent
researchers for study. Peer-reviewed research that concretely
substantiates the need for specific frameworks for data sharing will be
essential to ensuring academics can contribute to public policy efforts
to advance Internet security. 
Unexplored details include many that we directly address in
this paper: definition of incidents and their impact; data formats to
accommodate comparisons; disclosure controls technologies and
access policies to allow rigorous independent analyses. 
We hope this work can guide regulators to consider (and
facilitate) the role of academic research in informing and
leveraging reporting regulations that can advance scientific
understanding of the DDoS landscape.\footnote{A shorter term 
goal for industry transparency would
be for companies or a third party to publish and share historic
versions of industry reports, rather than only the most recent,
sometimes limited in distribution.}

\paragraph{Measurement of spoofing}
\label{subsec:measure_spoofing}
Spoofing persists as a key vector of DDoS attacks, and 
remediation of this vulnerability requires knowing which 
networks allow spoofing.  Attempts to measure source
address validation (SAV) have struggled with sustainability
for decades, \eg \cite{spoofer}. 
Efforts to internalize this negative externality -- 
``naming and shaming'', procurement guidelines, voluntary 
code of conducts -- have had limited 
impact \cite{luckieNetworkHygieneIncentives2019}.

We see two paths forward. First, a requirement for
transparency regarding SAV deployment, similar to other 
recent ISP transparency requirements \cite{fcc-rif,broadband-labels-2022}.
This direction would require a sustained operational
measurement infrastructure to support auditing
and impact assessments. 
Although long controversial, the RIPE Atlas measurement system
could support such measurements~\cite{aben_atlas_2016}.
Second, industry and governments could collaborate
to promote SAV measurement capability as a
default on end user equipment, a compromise on making SAV deployment
itself a default.


\paragraph{Availability measurement} \label{subsec:availability}
Breaches of confidentiality are now accepted as sufficiently
important to mandate reporting, but regulatory attention 
to requiring data on availability is in early 
stages.\footnote{In the UK, consumer ISPs must  reimburse customers £9.33 for each calendar day where the service is unavailable~\cite{noauthor_automatic_2023}, telecoms providers must notify Ofcom of availability incidents~\cite{ofcom_general_2003}, and banks must notify the Financial Conduct Authority~\cite{financial_conduct_authority_interpreting_2019}.}
A non-governmental approach could rely on financial sector
auditors, \eg PCI-DSS, to add availability to its auditing
framework.

\vspace{-2pt}

\section{Summary and Future Work}
\label{sec:conclusion}

We provided the first comprehensive, longitudinal view on direct-path and 
reflection amplification attacks, two dominant classes of DDoS~attacks
that threaten Internet infrastructure.
Our results document joint forces of academia and industry sharing 
data across institutional boundaries.
We included all datasets that were made available to us,
covering far more attacks than any other study.
Our approach to synthesizing datasets expanded beyond prior efforts in five dimensions: number of macroscopic datasets (ten), observation window (4.5 years vs months or even 3 years in previous works), multiple types of DDoS attacks, volume and usage restrictions of data (which required substantial cooperation across institutions in processing and normalizing data), and a new method to facilitate sharing by industry.
Beyond the synthesis of datasets, our work reinforced findings that previous studies have serious visibility limitations, which provide the strongest empirical grounding to date for regulatory framing to share data.

Our artifacts include a living compilation of facts from industry reports
characterizing DDoS phenomena in 2022-2023, and a mindmap taxonomy of
academic DDoS studies over the last few years (\autoref{appendix:artifacts}).
The next step is to motivate others who have DDoS data
to contribute to our effort to establish and maintain a holistic view 
of the DDoS ecosystem, following this blueprint.

\begin{acks}
We thank our shepherd and reviewers for helpful feedback.
We gratefully acknowledge all partner teams from industry and academia for sharing their data.
This work was partly supported by National Science Foundation grant CNS-2212241 and OAC-2319959, the German Research Foundation (DFG) within the project ReNO (\#511099228), and the German Federal Ministry of Education and Research (BMBF) within the project PRIMEnet.
\end{acks}

\label{lastpagebody}

\balance

\bibliographystyle{ACM-Reference-Format}
\bibliography{bibliography,DDoS-reports}

\appendix

\appendix

\section{Ethics}
\label{appendix:ethics}

We considered the following ethical concerns.

\paragraph{Harmful data collection}
Our analysis is based on data sets extensively used in prior work, not of itself a justification~\cite{Thomas2017a}, however, we do no additional harm. 
DDoS \textit{honeypots} deploy safeguards to avoid participating in attacks while still collecting as much information about attackers as possible (often reducing attack traffic).
\textit{Network telescopes} are purely passive instruments and do not participate in any activity.
\textit{On path flow monitoring} data from IXPs and DDoS mitigation providers is already collected for operational purposes, which are supported by the insights from research reusing the data.
We worked with a wide range of stakeholders to minimize unanticipated risks~\cite{Thomas2017a}.

\paragraph{Leaking personal information}
Our analysis focuses on aggregate trends and correlations.
We do not reveal individual IP addresses, or any personal information.
Correlations with customer information was done by a party already trusted by the customer without leaking information into the collected data set.
Some datasets are shared in a controlled way with researchers by their original sources~\cite{CCCCDataAgreements} enabling reproducibility while maintaining safeguards.

\paragraph{Data anonymization}
The data from industry does not include any personal identifying information (PII) as we received attack counts of different granularity that we aggregate to weeks.
While we have access to attack event data for the research observatories, we do not include any PII information in the paper.
As such, no anonymization was required to present our results.

\section{Artifacts}
\label{appendix:artifacts}

We contribute two artifacts that categorize related work on DDoS attacks. We believe this work can serve the community as a reference.

\paragraph{Industry: DDoS report survey}
Throughout the DDoS industry, companies report on their observation on DDoS attacks.
We present a deep dive into reports released around 2022 in \autoref{sec:industry-reports} alongside supplemental materials in \autoref{appendix:industry-reports} and online~\cite{ddos-industry-github}.

\paragraph{Academia: related work taxonomy}
Work on DDoS in research has been extensive.
\autoref{sec:related-work} presents a condensed overview that is systematically categorized based on research field and data sets in \autoref{appendix:relwork}.

\section{Taxonomy ``Mindmap'' of DDoS Literature}
\label{appendix:relwork}

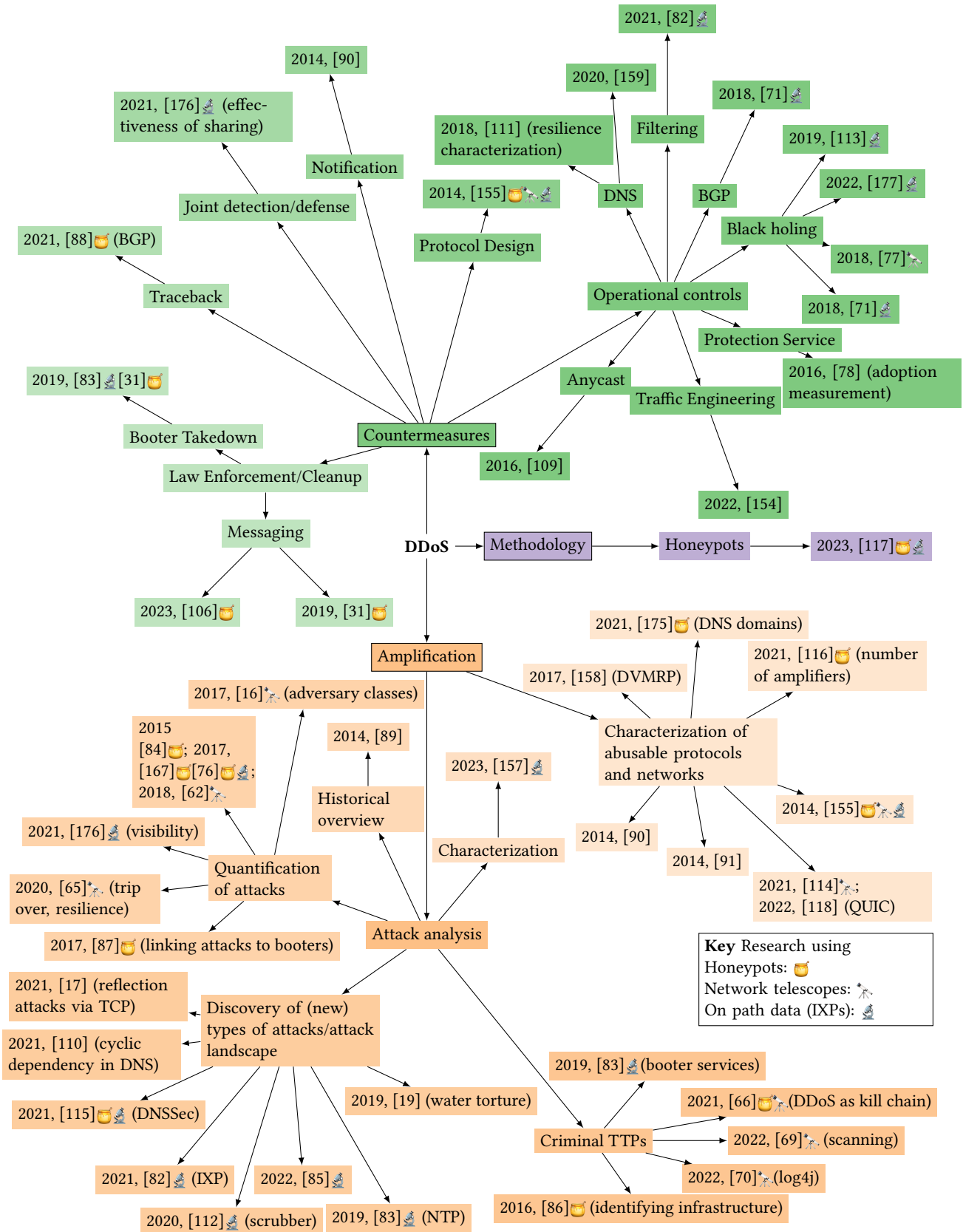
\begin{figure*}
\definecolor{countermeasures}{HTML}{7fc97f}
\definecolor{methodology}{HTML}{beaed4}
\definecolor{amplification}{HTML}{fdc086}
\begin{tikzpicture}[%
	level 1/.style={level distance=2cm,sibling angle=90},
	level 2/.style={level distance=5cm,sibling angle=45},
	level 3/.style={level distance=2cm,sibling angle=60},
	level 4/.style={level distance=2cm,sibling angle=90},
	level 5/.style={level distance=2cm,sibling angle=90},
	edge from parent/.style={draw,-latex}]%
\node{\textbf{DDoS}}
	child[grow=90, every node/.style={fill=countermeasures}] { node[draw] {Countermeasures}
		child[grow=105, every node/.style={fill=countermeasures!80}] { node {Notification}
			child { node {2014, \cite{kuhrer2014exit}}}
		}
		child[grow=75, level distance=3.5cm, every node/.style={fill=countermeasures!90}] { node {Protocol Design}
			child[level distance=1cm] { node {2014, \cite{r-ahrnp-14}\honeypot\telescope\dpi}}
		}
		child[grow=30] { node {Operational controls}
			child[grow=90, level distance=3cm] { node {Filtering}
				child { node {2021, \cite{kdh-dndip-21}\dpi}}
			}
			child[grow=65] { node {BGP}
				child { node {2018, \cite{hnjds-pfdms-18}\dpi}}
			}
			child[grow=115] { node {DNS}
				child { node {2020, \cite{sbkms-adpaa-20}}}
				child[grow=150] { node[text width=3cm] {2018, \cite{mhmsd-wdbdd-18} (resilience characterization)}}
			}
			child[grow=20,yshift=0.5cm] { node {Black holing}
				child[grow=25] { node {2022, \cite{ixpscrubber}\dpi}}
				child[grow=55] { node {2019, \cite{nawrocki2019down}\dpi}}
				child[grow=-45] { node {2018, \cite{hnjds-pfdms-18}\dpi}}
				child[grow=-15] { node {2018, \cite{jonker2018-blackholing}\telescope}}
			}
			child[grow=-70] { node {Traffic Engineering}
				child { node {2022, \cite{rbch-aanpf-22}}}
			}
			child[grow=-130] { node {Anycast}
				child { node {2016, \cite{mshvm-adenr-16}}}
			}
			child[grow=-23] { node {Protection Service}
				child { node[text width=3cm] {2016, \cite{jonker2016protection} (adoption measurement)}}
			}
		}
		child[grow=125, every node/.style={fill=countermeasures!70}] { node {Joint detection/defense}
			child { node[text width=3cm] {2021,~\cite{DXP}\dpi{} (effectiveness of sharing)}}
		}
		child[grow=150, every node/.style={fill=countermeasures!60}] { node {Traceback}
			child { node {2021, \cite{kr-babta-21}\honeypot{} (BGP)}}
		}
		child[grow=195,level distance=3cm, every node/.style={fill=countermeasures!50}] { node {Law Enforcement/Cleanup}
			child[grow=150,level distance=1.5cm] { node {Booter Takedown}
				child { node {2019, \cite{kopp2019hide}\dpi\cite{BooterShutdown}\honeypot}}
			}
			child[grow=-90,level distance=1cm] { node {Messaging}
				child[grow=-45] { node {2019, \cite{BooterShutdown}\honeypot}}
				child[grow=-135] { node {2023, \cite{moneva_effect_2023}\honeypot}}
			}
		}
	}
	child[grow=-90, every node/.style={fill=amplification}] { node[draw] {Amplification}
		child[grow=-20, every node/.style={fill=amplification!40}] { node[text width=3cm] {Characterization of abusable protocols and networks}
			child[grow=-130] { node {2014, \cite{kuhrer2014exit}}}
			child[grow=-80] { node {2014, \cite{khrh-hhatr-14}}}
			child[level distance=3.5cm] { node[text width=3cm] {2021,~\cite{nawrocki2021quic}\telescope; 2022,~\cite{nthms-ibtcq-22} (QUIC)}}
			child[level distance=3cm] { node {2014, \cite{r-ahrnp-14}\honeypot\telescope\dpi}}
			child[grow=30,level distance=3.1cm] { node[text width=3cm] {2021,~\cite{nawrocki2021transparent}\honeypot{} (number of amplifiers)}}
			child[grow=85,level distance=2.3cm] { node {2021, \cite{tkjrr-amadp-21}\honeypot{} (DNS domains)}}
			child[grow=138] { node {2017, \cite{akpa-opai-17} (DVMRP)}}
		}
		child[grow=-90] { node {Attack analysis}
			child[grow=50, every node/.style={fill=amplification!50}] { node {Characterization}
				child[grow=90, level distance=1.5cm] { node {2023, \cite{simra2023ddos2vec}\dpi}}
			}
			child[grow=115,level distance=2.5cm, every node/.style={fill=amplification!60}] { node[text width=1.8cm] {Historical overview}
				child[grow=90, level distance=1.3cm] { node {2014, \cite{kk-nr-14}}}
			}
			child[grow=160,level distance=3cm, every node/.style={fill=amplification!70}] { node[text width=2cm] {Quantification of attacks}
				child[xshift=1.5cm,yshift=-0.2cm] { node {2017, \cite{bgd-qsdai-17}\telescope{} (adversary classes)}}
				child { node[text width=2cm] {2015 \cite{kramer2015amppot}\honeypot; 2017, \cite{thomas2017thousand}\honeypot\cite{jkkrs-mtuam-17}\honeypot\dpi{}; 2018, \cite{gd-wtmc-2018}\telescope}}
				child[level distance=3cm,yshift=-0.2cm] { node {2021, \cite{DXP}\dpi{} (visibility)}}
				child[level distance=3cm] { node[text width=2.5cm] {2020, \cite{gd-qtsdr-20}\telescope{} (trip over, resilience)}}
				child[yshift=0.9cm, xshift=1.5cm] { node {2017, \cite{dbpa-ladab-17}\honeypot{} (linking attacks to booters)}}
			} 
			child[grow=-145,level distance=3cm, every node/.style={fill=amplification!80}] { node[text width=3cm] {Discovery of (new) types of attacks/attack landscape}
				child[grow=-25, level distance=3cm] { node {2019, \cite{b-ccdda-19} (water torture)}}
				child[grow=-60, level distance=3.9cm] { node {2019, \cite{kopp2019hide}\dpi{} (NTP)}}
				child[grow=-107, level distance=3.6cm] { node {2020, \cite{mhsbw-idmls-20}\dpi{} (scrubber)}}
				child[grow=-130, level distance=3.5cm] { node {2021, \cite{kdh-dndip-21}\dpi{} (IXP)}}
				child[grow=-155, level distance=3.6cm] { node {2021, \cite{nawrocki2021ecosystem}\honeypot\dpi{} (DNSSec)}}
				child[grow=-173,level distance=3.6cm] { node[text width=3cm] {2021, \cite{mchh-temvd-21} (cyclic dependency in DNS)}}
				child[grow=-190, level distance=3.5cm] { node[text width=3cm] {2021, \cite{bafhw-wmtra-21} (reflection attacks via TCP)}}
				child[grow=-85, level distance=2.7cm] { node {2022, \cite{k-dtlr-22}\dpi}}
			}
			child[level distance=3.7cm, every node/.style={fill=amplification!90}] { node {Criminal TTPs}
				child[grow=55,yshift=-0.3cm] { node {2019, \cite{kopp2019hide}\dpi (booter services)}}
				child[grow=20,xshift=2cm] { node {2021, \cite{griffioen2021adversarial}\honeypot\telescope (DDoS as kill chain)}}
				child[grow=0,xshift=2cm] { node {2022, \cite{hnkds-sunws-22}\telescope{} (scanning)}}
				child[grow=-20,xshift=1cm] { node {2022, \cite{hnsw-rvmls-22}\telescope (log4j)}}
				child[grow=-55, level distance=1.5cm] { node {2016, \cite{kbr-issai-16}\honeypot{} (identifying infrastructure)}}
			}
		}
	}
	child[grow=0, every node/.style={fill=methodology}] { node[draw] {Methodology}
		child[xshift=-2cm] { node {Honeypots}
			child[xshift=1cm] { node {2023, \cite{nkkhs-advmd-23}\honeypot\dpi}}
		}
	}
;
\node (key) [draw,text width=4cm] at (7,-7.8) {\textbf{Key} Research using\\ Honeypots: \honeypot\\ Network telescopes: \telescope\\ On path data (IXPs): \dpi} ;
\end{tikzpicture}
\caption{Taxonomy of related work in DDoS. }
\label{fig:related-work-taxonomy}
\end{figure*}

We provide a graphical (``mindmap'') taxonomy of the extensive relevant
literature in the last few years on the two dominant classes of attacks 
that we study (Direct-Path and Reflection/Amplification attacks).
Figure~\ref{fig:related-work-taxonomy} illustrates the many dimensions
of the problem that have received focused attention.  The figure is not 
exhaustive -- there are hundreds of other papers on DDoS, which
we expect will mostly fall within the themes included within this taxonomy.
Our takeaway from this analysis is that while there is an abundance of
research activity related to DDoS, there is little effort to find a
convergent position on how effective current defenses are against the
threat.

\section{Honeypot: NewKid}
\label{appendix:newkid}

\begin{figure}
  \centering
  \includegraphics[width=1.0\linewidth]{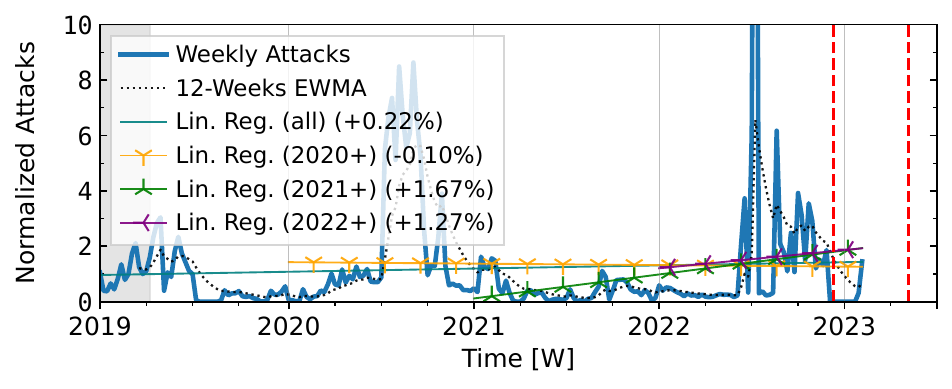}
  \caption{Normalized attack trends observed by of the NewKid honeypot. The peak in mid 2022 rises up to 33.}
  \label{fig:normalized_attack_trends:newkid} 
\end{figure}

\autoref{fig:normalized_attack_trends:newkid} shows attack counts 
observed by the NewKid honeypot, which consists of a single sensor
whose observations are erratic, although still reasonably consistent
with the macroscopic data sets we analyzed.
For example, periods of high attacks around mid-2020 and
mid-2022 match peaks in attacks observed by
Hopscotch~\autoref{fig:normalized_attack_trends:ccc} and
AmpPot~\autoref{fig:normalized_attack_trends:amppot}.

\section{Summary of Industry Reports}
\label{appendix:industry-reports}

\begin{table}
	\caption{List of documents assessed.}
	\label{tab:report_summary}
	\centering
	\begin{tabular}{lll}
	\toprule
	Company & Included & Omitted \\
	\midrule
	A10              & \cite{a102022A10Networks2022}                                             & \\
	Akamai           & \cite{akamaiDDoSAttacks20222023, akamaiRelentlessEvolutionDDoS2022}       & \\
	Alibaba Cloud    &                                                                           & \cite{alibabacloudDDoSAttackStatistics2021}                                                           \\
	AWS              &                                                                           & \cite{awsShieldThreat2021}                                                                            \\
	Arelion          & \cite{arelionArelionDDoSThreat2023}                                       & \\
	Cloudflare       & \cite{cloudflareDDoSAttackTrendsQ42022}                                   & \cite{cloudflareDDoSTrends2023,cloudflareDDoSAttackTrendsQ12022,cloudflareDDoSAttackTrendsQ22022,CloudflareDDoSThreat2022} \\
	Comcast          & \cite{comcast2023ComcastBusiness2023}                                     & \cite{comcastBusinessDDoS2021} \\
	Corero           & \cite{corero2023DDoSThreat2023}                                           & \cite{corero2023DDoSAttacksEvolved, corero2023ShiftingLandscapeDDoSAttacks} \\
	Crowdstrike      &                                                                           & \cite{crowdstrike2023GlobalThreat2023} \\
	DDoS-Guard       & \cite{ddos-guardDDoSAttackTrends2023, ddos-guardDDoSGuardAnalyticalReport2023} & \\
	F5               & \cite{f5F5DDoSAttack2023}                                                 & \\
	Fastly           &                                                                   			& \cite{fastlyCyberThreatInsights2023, fastlyWhatDDoSAttack2023} \\
	Fortinet         &                                                                   			& \cite{fortinetGlobalThreatLandscape2023} \\
	Huawei           & \cite{huaweiGlobalDDoSAttack}                                             & \\
	Imperva          & \cite{impervaImpervaGlobalDDoS2023}                                       & \\
	Kapersky         & \cite{kasperskyKaperskyDDoSAttacks2022}                                   & \cite{kasperskyKasperskyDDoSReport2022, kasperskyKaperskyDDoSAttacks2022a}          \\
	LINK11           & \cite{link11LINK11DDOSREPORT20222023}                                     & \\
	Lumen            & \cite{lumenLumenQuarterlyDDoSQ42022}                                      & \cite{lumenLumenQuarterlyDDoSQ32022, labsTrackingUDPReflectors2021}                                   \\
	Microsoft        & \cite{microsoftazurenetworksecurityteam2022ReviewDDoS2023}                & \\
	NBIP             & \cite{nbipDDoSAttackFiguresFourthQ2023b}                                  & \cite{nbipDDoSAttackFigures2023, nbipDDoSAttackFigures2023a}                                          \\
	Netscout         & \cite{netscout5thAnniversaryDDoS2023}                                     & \cite{netscoutNETSCOUTThreatIntelligence2021, netscoutNETSCOUTDDoSAttack2023} \\
	NexusGuard       & \cite{nexusguardDDoSStatisticalReport2023}                                & \cite{nexusguardDDoSStatisticalReport2023a}                                                           \\
	Nokia            & \cite{nokiaNokiaThreatIntelligence2023}                                   & \cite{nokiaNokiaDeepfieldNetwork2022, craiglabovitzTracingDDoSEndtoEnd2021}                           \\
	NSFocus          & \cite{nsfocus2022GlobalDDoS2023}                                          & \cite{nokiaChangingDDoSThreat2022}                                                                    \\
	Palo Alto        &                                                                           & \cite{paloaltoUnit42INCIDENT}                                                                         \\
	Qrator           & \cite{qratorQ42022DDoS23}                                                 & \cite{qratorQ12022DDoS2022,qratorQ22022DDoS2022,qratorQ32022DDoS2022}\\
	Radware          & \cite{radwareRadwareGlobalThreat2023}                                     & \\
	RioRey           &                                                                           & \cite{rioreyRioReyTaxonomyDDoS2015}                                                                   \\
	Splunk           &                                                                           & \cite{splunkDenialofServiceAttacksHistory2023}                                                        \\
	Zayo             & \cite{zayoProtectingYourBusiness2023}                                     & \cite{zayoLookRecentDDoS2022}                                                                     \\
\bottomrule
	\end{tabular}
\end{table}

\autoref{tab:report_summary} lists vendors that published reports,
and highlights the reports we discuss in this paper.
We created an extensive table \cite{ddos-industry-github} containing 
information we extracted from industry reports.
The table is a living artifact
of this work; we invite interested community members (including
report authors!) to expand the table with additional reports, 
or annotations of existing reports.
We imagine this table as a potential 
tool to support pursuit of some measure of 
\emph{community consensus} regarding
the DDoS ecosystem.  Further, we hope it provides some incentive for
industry players to engage in conversation to elucidate their results.
An accompanying git repository \cite{ddos-industry-github} contains 
the table in PDF and other formats and all the files used 
as sources
for our analysis.  

The columns in this comprehensive table \cite{ddos-industry-github}
include: vendor,
Document Description,
title,
format,
period of analysis,
attack counts,
reflection/amplification vectors,
direct path vectors,
other vectors,
changes during 2022,
attack duration/size,
attack intensity,
carpet bombing,
multi-vector,
target,
vantage points/sources.

\begin{figure}
  \centering
  \includegraphics[trim={2.8cm 0 0 0},clip,width=1.0\linewidth]{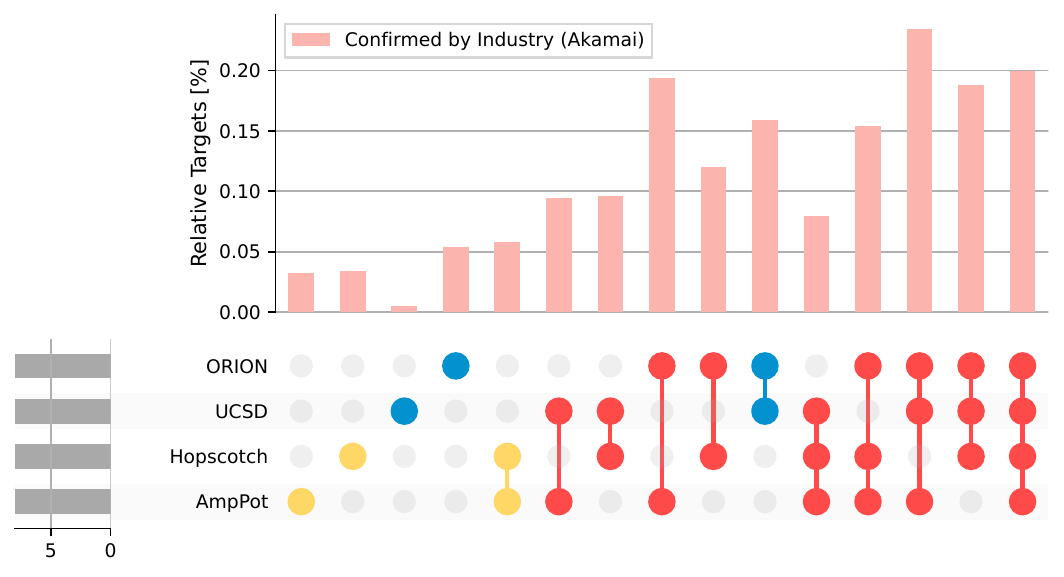}
  \caption{Relative share of targets observed by academic observatories and confirmed by industrial baseline data from \textit{Akamai}.}
  \label{fig:industry:akamai}
\end{figure}

\section{Quarterly Correlations}
\label{appendix:correlation:quarterly}

\begin{figure*}
  \centering
  \includegraphics[width=1.0\linewidth]{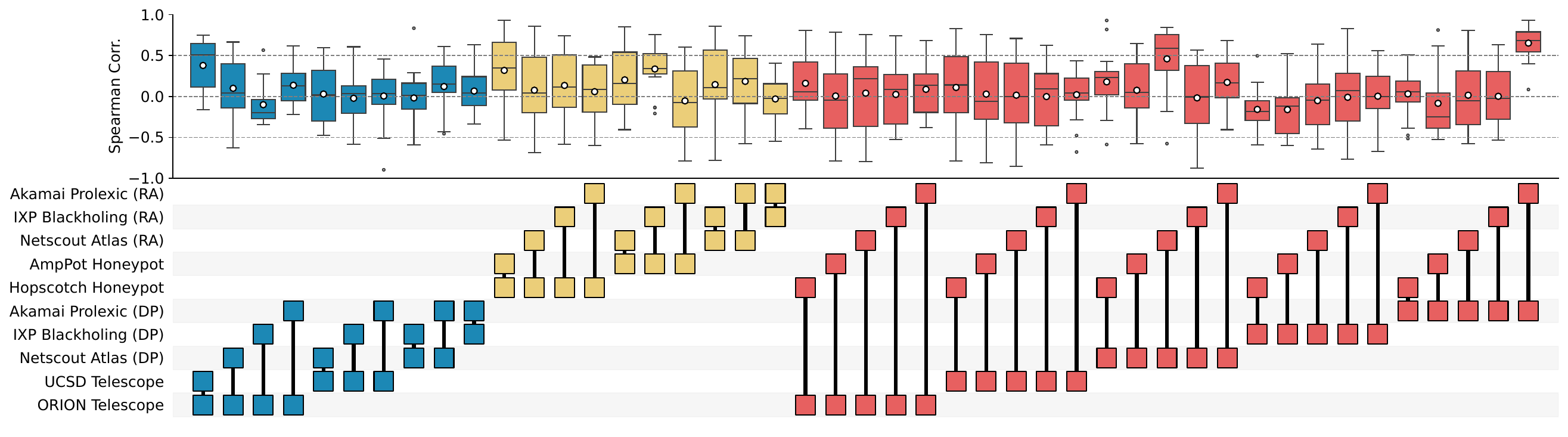}
  \caption{Quarterly pair-wise correlations across observatories from academia and industry. Vertical bars mark the median and the circles the mean.}
  \label{fig:correlation:quarterly}
\end{figure*}

\autoref{subsec:trends_across_attack_types} discusses the correlations of long-term trends and summarizes them in \autoref{fig:vp_correlation:spearman}.
\autoref{fig:correlation:quarterly} shows a coarse-grained view on pair-wise correlations.
Each box summarizes quarterly Spearman correlations (18 values over 4.5 years) for the pair of observatories marked in the matrix below.
Vertical bars mark the median and the circles the mean.
The coloring scheme matches the UpSet plots in \autoref{sec:targets}, \textit{blue}: direct-path observatories, \textit{yellow}: reflection-amplification observatories, and \textit{red}: correlations across attack types.

Most correlations are not stable across quarters, \ie the whiskers cover a large part of the possible correlations (-1 to +1).
Two exceptions are the pair AmpPot \& IXP Blackholing (RA) and the pair Akamai Prolexic (DP) \& Akamai Prolexic (RA), which only have outliers in the anti-correlation range, but generally show a high correlation with each other.

The whiskers for box that correlate observatories of the same attack type tend to have less variance.
While they reach less into anti-correlations overall, they often mix periods of correlation and anti-correlation.
The boxes for mixed correlations tend to be focused more around 0, although the DDoS mitigation services (Netscout, Akamai) both show higher correlations among their own two time series.

\section{Akamai Target Overlap}
\label{appendix:targets:akamai}

\autoref{fig:industry:akamai} shows how many of the DDoS target tuples (date, IP address) observed by academia (\autoref{fig:targets:overview}) were also observed by Akamai.
Similar to Netscout (\autoref{fig:industry:netscout}) the shares were higher for attacks visible at multiple observatories.
These are likely highly visible attacks.

\section{Overlap of Targeted ASes}
\label{appendix:overlap-asn}

\autoref{tab:cloud_providers} shows the top 10 ASes observed as DDoS targets (date, IP) at ORION, UCSD, Hopscotch, and AmpPot (\autoref{subsec:Attackers_Usually_Chose_One_Attack_Type}). All are labeled as hosting ASes except for Microsoft (business), China Unicom (ISP), and Alibaba (business).

\begin{table}
\centering
\begin{tabular}{rlrrrl}
\toprule
Rank & Provider & ASN & Tuples & Share \\
\midrule
1 & OVH & 16276 & \num{28769} & 18.80\% \\
2 & Hetzner & 24940 & \num{7872} & 5.14\%  \\
3 & Amazon & 16509 & \num{4123} & 2.69\%  \\
4 & Microsoft & 8075 & \num{3125} & 2.04\%  \\
5 & Google & 396982 & \num{2898} & 1.89\%  \\
6 & Cloudflare & 13335 & \num{2427} & 1.59\% \\
7 & China Unicom & 4837 & \num{2421} & 1.58\% \\
8 & Digitalocean & 14061 & \num{2081} & 1.36\%  \\
9 & Nuclearfallout & 14586 & \num{1885} & 1.23\% \\
10 & Alibaba & 37963 & \num{1847} & 1.21\% \\
\bottomrule
\end{tabular}
\caption{Top 10 ASes that were observed in all our four research observatories (ORION, UCSD, Hopscotch, AmpPot).}
\label{tab:cloud_providers}
\end{table}

\section{Detecting Carpet Bombing/Prefix Attacks in Honeypot Data}
\label{appendix:prefix}

Spreading one attack over many IP addresses (variously called ``carpet bombing'', ``prefix attacks'' or ``horizontal attacks'')
makes it more difficult to block, or even identify as a single attack.
Honeypots in particular struggle with this challenge because
they are generally comprised of many IP addresses, and attackers
deploy a range of strategies to select target IP addresses
(randomized, sequential, bursty) and so honeypots may not observe even one packet for every targeted address.
Our approach to inferring (effectively reconstructing) such attacks
builds on prior work \cite[Appendix A-C]{thomas2017thousand} to 
aggregate attacks within the same IP prefix. Our method essentially
finds the longest BGP-routed prefix (from /11 to /28) that covers the attack.

Our approach does not aggregate attacks that span multiple IP address block allocations (from Regional Internet Registry (RIR) data).
This means that when an attacker targets many blocks of IP addresses that are allocated to the same AS, because they are targeting an ISP rather than a customer, this is recorded as many attacks rather than a single attack.
Such attacks targeted Brazil using SSDP in mid-2022 causing the spikes 
in~\autoref{fig:normalized_attack_trends:ccc} and~\ref{fig:normalized_attack_trends:amppot}.  Algorithms that detect attacks targeting entire ASes might 
remove this noise.  We provided the details of our algorithm in a 
script we shared with the Cambridge Cybercrime Center for use in 
their honeypot infrastructure; they will make it available to
researchers on request~\cite{CCCCDataAgreements}.

\section{RSDoS Inference from Network Telescopes}
\label{sec:app-rsdos-infer}

We provide implementation details and references for the RSDoS analysis in \autoref{sec:long-term-trends}.
We used CAIDA's Corsaro tools~\cite{corsaro-github}, which is based on~\cite{2006-moore-bd}.
Corsaro uses a \textit{flow identifier} to group packets into flows.
A packet \textit{threshold} and \textit{timeout} discern which of these flows are part of an attack and when this attack stops.

\begin{enumerate}
\item {\bf Flow identifier}: The tuple (protocol, source IP) identifies a flow. In code, packets are matched in two steps: \one the protocol selects a hashmap, \two the source IP identifies the flow within it. Source and destination ports are aggregated as part of the data rather than part of the key\footnote{\url{https://github.com/CAIDA/corsaro3/blob/master/libcorsaro/plugins/corsaro\_dos.c\#L1014}}.

\item {\bf Threshold}: To be considered an attack a flow must have a minimum of 25 packets from a single source IP and last for 60 seconds. Additionally, an attack flow must (at one point) meet a packet rate of at least 30 packets across a 60-second window, which slides every 10 seconds\footnote{\url{https://github.com/CAIDA/corsaro3/wiki/DoS-Plugin\#configuration}}.

\item {\bf Timeout}: Corsaro counts packets in intervals of 300 seconds. After an interval with no new packets an attack flow is finished.%
\footnote{\url{https://github.com/CAIDA/corsaro3/blob/master/libcorsaro/plugins/corsaro\_dos.c\#L1251}}

\end{enumerate}

Once both thresholds (packet count, packet rate) have been met (at any point in the flow) that flow counts as an attack for the rest of its lifetime.
Any number of packets is enough to maintain it until the flow times out because no new packets arrive.
(We are not sure why the authors evolved the code this way.) 
The default values mentioned are explicitly set in the Corsaro config file used in the analysis (\url{https://github.com/CAIDA/corsaro3/blob/master/libcorsaro/plugins/corsaro\_dos.c\#L1265)}).

\label{lastpage}

\end{document}